\documentclass[useAMS,usenatbib]{mn2e}

\usepackage{enumerate}
\usepackage{gensymb}
\usepackage{ textcomp }
\usepackage{graphicx}
\usepackage{caption}
\usepackage{subcaption}

\title[The giant lobes of Centaurus A observed at 118 MHz with the Murchison Widefield Array]{The giant lobes of Centaurus A observed at 118 MHz with the Murchison Widefield Array}

\author[B. McKinley et al.]
{B.~McKinley,$^{1,2}$\thanks{E-mail:ben@mso.anu.edu.au}
F.~Briggs,$^{1,2}$
B.~M.~Gaensler,$^{3,2}$
I.~J.~Feain,$^4$
G.~Bernardi,$^{5,6}$
\newauthor
R.~B.~Wayth,$^{7,2}$
M.~Johnston-Hollitt,$^8$
A.~R.~Offringa,$^{1,2}$ 
W.~Arcus,$^7$
D.~G.~Barnes,$^9$
\newauthor
J.~D.~Bowman,$^{10}$
J.~D.~Bunton,$^4$
R.~J.~Cappallo,$^{11}$
B.~E.~Corey,$^{11}$
A.~Deshpande,$^{12}$
\newauthor
L.~deSouza,$^{4,3}$
D.~Emrich,$^7$
R.~Goeke,$^{13}$
L.~J.~Greenhill,$^5$
B.~J.~Hazelton,$^{14}$
D.~Herne,$^7$
\newauthor
J.~N.~Hewitt,$^{13}$ 
D.~L.~Kaplan,$^{15}$
J.~C.~Kasper,$^5$
B.~B.~Kincaid,$^{11}$
R.~Koenig,$^4$
\newauthor
E.~Kratzenberg,$^{11}$
C.~J.~Lonsdale,$^{11}$
M.~J.~Lynch,$^7$
S.~R.~McWhirter,$^{11}$
D.~A.~Mitchell,$^{16,2}$
\newauthor
M.~F.~Morales,$^{14}$
E.~Morgan,$^{13}$
D.~Oberoi,$^{17}$
S.~M.~Ord,$^{7,2}$
J.~Pathikulangara,$^4$
\newauthor
T.~Prabu,$^{12}$
R.~A.~Remillard,$^{13}$
A.~E.~E.~Rogers,$^{11}$
A.~Roshi,$^{12}$
J.~E.~Salah,$^{11}$
\newauthor
R.~J.~Sault,$^{16}$
N.~Udaya~Shankar,$^{12}$
K.~S.~Srivani,$^{12}$
J.~Stevens,$^{4,18}$
R.~Subrahmanyan,$^{12,2}$
\newauthor
S.~J.~Tingay,$^{7,2}$
M.~Waterson,$^{7,1}$
R.~L.~Webster,$^{16,2}$
A.~R.~Whitney,$^{11}$
A.~Williams,$^{7}$
\newauthor
C.~L.~Williams,$^{13}$
J.~S.~B.~Wyithe,$^{16,2}$
\\
$^{1}$Research School of Astronomy and Astrophysics, Australian National University, Canberra, Australia\\
$^{2}$ARC Centre of Excellence for All-sky Astrophysics (CAASTRO)\\
$^{3}$Sydney Institute for Astronomy, School of Physics, The University of Sydney, NSW 2006, Australia\\
$^{4}$CSIRO Astronomy and Space Science, Marsfield NSW, Australia\\
$^{5}$Harvard-Smithsonian Center for Astrophysics, Cambridge MA, USA\\
$^{6}$Square Kilometre Array South Africa (SKA SA), Cape Town, South Africa\\
$^{7}$International Centre for Radio Astronomy Research, Curtin University, Bentley WA, Australia\\
$^{8}$School of Chemical \& Physical Sciences, Victoria University of Wellington, New Zealand\\
$^{9}$Monash e-Research Centre, Monash University, Clayton VIC, Australia\\
$^{10}$School of Earth and Space Exploration, Arizona State University, Tempe AZ, USA\\
$^{11}$MIT Haystack Observatory, Westford MA, USA\\
$^{12}$Raman Research Institute, Bangalore, India\\
$^{13}$Kavli Institute for Astrophysics and Space Research, Massachusetts Institute of Technology, Cambridge MA, USA\\
$^{14}$Department of Physics, University of Washington, Seattle WA, USA\\
$^{15}$Department of Physics, University of Wisconsin--Milwaukee, Milwaukee WI, USA\\
$^{16}$School of Physics, The University of Melbourne, Parkville VIC, Australia\\
$^{17}$National Centre for Radio Astrophysics, Tata Institute for Fundamental Research, Pune, India\\
$^{18}$School of Mathematics and Physics, University of Tasmania, Hobart, Australia\\
}
\begin{document}

\date{Accepted 1988 December 15. Received 1988 December 14; in original form 1988 October 11}

\pagerange{\pageref{firstpage}--\pageref{lastpage}} \pubyear{2002}

\maketitle

\label{firstpage}

\begin{abstract}
We present new wide-field observations of Centaurus A (Cen~A) and the surrounding region at 118~MHz with the Murchison Widefield Array (MWA) 32-tile prototype, with which we investigate the spectral-index distribution of Cen~A's giant radio lobes. We compare our images to 1.4~GHz maps of Cen~A and compute spectral indices using temperature-temperature plots and spectral tomography. We find that the morphologies at 118~MHz and 1.4~GHz match very closely apart from an extra peak in the southern lobe at 118~MHz, which provides tentative evidence for the existence of a southern counterpart to the northern middle lobe of Cen~A. Our spatially-averaged spectral indices for both the northern and southern lobes are consistent with previous analyses, however we find significant spatial variation of the spectra across the extent of each lobe. Both the spectral-index distribution and the morphology at low radio frequencies support a scenario of multiple outbursts of activity from the central engine. Our results are consistent with inverse-Compton modelling of radio and gamma-ray data that supports a value for the lobe age of between 10 and 80~Myr.\\
\end{abstract}


\section{Introduction}

Radio galaxies consist of two or more extended regions of magnetised plasma and synchrotron-emitting, relativistic charged particles, emanating from the active galactic nucleus (AGN) of a host galaxy \citep{begelman}. These radio lobes interact with the intergalactic medium and/or intracluster medium and grow in size over time, due to either a quasi-continuous injection of particles from the jets or possibly separate episodic outbursts \citep{ravi,morganti}. At GHz frequencies, the emission spectra of radio galaxies, within the observing band of a typical radio telescope, are often well described by a simple power law of the form $S\propto \nu^{\alpha}$, as a function of frequency, $\nu$, where $S$ is the observed flux density and $\alpha$ is the spectral index. The power law spectrum results from the power law energy distribution of relativistic particles producing the observed synchrotron emission. Inverse Compton emission may also be observed in radio galaxies when relativistic particles upscatter low-energy background photons, the most abundant of which are from the cosmic microwave background, to higher energies. Combining X-ray and gamma-ray data with radio data can therefore be used to constrain models of synchrotron and inverse Compton emission to estimate the physical properties of radio galaxies.

One of the most well-known and well-studied radio galaxies is Centaurus A (Cen~A). This Fanaroff \& Riley (FRI, \citealt{FR}) radio source associated with the elliptical galaxy NGC5128 was first discovered by \citet{bolton} at 100~MHz, the frequency we return to here in this paper. Cen~A lies at a distance of a mere $3.8\pm0.1$~Mpc \citep{harris}, making it the closest radio galaxy by a large margin and providing an ideal laboratory for studying the properties and particle acceleration histories of radio galaxies. Cen~A has been studied intensely over a wide range of wavelengths and physical scales (see \citealt{israel} for a review), but there is still much to be learned from this complex, nearby object. 


The giant lobes of Cen~A have a distinctive morphology with multiple peaks and orientations that change with distance from the core. They comprise of a pair of inner lobes emanating from the core of the galaxy with a combined extent of $\sim$11 kpc, aligned at a position angle of $55\degr$ anti-clockwise from the north-south axis \citep{schreier,vla}. Further from the core, the northern inner lobe is followed by a `northern middle lobe' (NML, \citealt{morganti}), which is the brightest part of the northern lobe situated  $\sim$30 kpc from the core and orientated at a position angle of $45\degr$. No southern counterpart to the NML has so far been detected in total intensity measurements \citep{feain2011}, but \citet{osullivan} have reported the existence of a region of high fractional polarisation in the southern lobe that could be the oppositely directed counterpart to the NML. Finally, there are the giant outer lobes, which have a combined extent of $\sim$500 kpc. The northern outer lobe has a position angle of between $0\degr$ and $20\degr$  closest to the core, and then has an abrupt change in position angle at approximately RA (J2000) 13\textsuperscript{h}25\textsuperscript{m}00\textsuperscript{s}, Dec (J2000) $-40\degr58\arcmin00\arcsec$ \citep{feain2011,haynes,cooper1965}. The southern outer lobe also experiences a change in position angle, but one that is more gradual. 

To produce the complex structure of Cen A, current models invoke a combination of separate, multiple outbursts of nuclear activity \citep{morganti}, combined with interaction with the ambient medium \citep{wats,burns} and a precessing central engine \citep{haynes}. An alternative explanation is that the jets interact with rotating gaseous shells (identified in optical observations by \citealt{malin}, in 21 cm emission by \citealt{HIshells}, and in the far-infrared by \citealt{stickel}) surrounding the host galaxy \citep{gopal1984,gopal2010}. Further observational and theoretical work is required to determine the origins and evolutionary history of the giant lobes.

A new generation of low-frequency radio interferometers is beginning to produce scientific results and can provide more information on the shape of the radio spectrum of radio galaxies such as Cen~A with reasonable angular resolution. These instruments, including the Low Frequency Array (LOFAR, \citealt{LOFAR}), the Precision Array for Probing the Epoch of Reionisation (PAPER, \citealt{PAPER}) and the Murchison Widefield Array (MWA, \citealt{tingay,bowman2013,lonsdale}), aim to detect the faint redshifted 21-cm spectral line signal from the epoch of reionisation (EoR). They have been designed to have the wide fields of view and short baselines required to detect the EoR signal, which is expected to have characteristic angular scales up to tens of arcmin (\citealt{pritchard_loeb,moraleswyithe2010,furlanetto2006}), as well as the longer baselines that provide the angular resolution required for accurate instrument calibration. Therefore, these types of instruments are also ideally suited to measuring the low-frequency properties of extended objects such as Cen~A.

In this paper we present a spatially resolved radio-frequency spectral-index study of the giant lobes of Cen~A using data from the MWA 32~Tile prototype (MWA 32T) at 118~MHz and Parkes data at 1.4 GHz  \citep{osullivan}. In Section 2, we discuss previous radio observations and spectral analyses of the giant lobes. In Section 3, we provide the details of the MWA observations and data reduction. In Section 4, we present the results of the MWA observations and our analysis of the spatial and spectral properties of the giant lobes between 118~MHz and 1.4~GHz. In Section 5, we discuss the results in relation to current models of the origin and evolution of the giant lobes.

\section[]{Previous observations and analyses of the giant lobes}

Cen~A has a large angular extent of $4\degr\times8\degr$, making observations of the \emph{entire} radio source difficult, especially when attempting to obtain images with reasonable angular resolution. Studies of the entire source at radio frequencies include full polarisation maps obtained by \citet{cooper1965} at 406, 960, 1410, 2650 and 5000~MHz using the Parkes Radio Telescope (with resolutions ranging from 48 to 4 arcmin depending on frequency) and polarisation and total intensity maps of the full source at 4.75~GHz and 4-arcmin resolution produced by \citet{junkes}. More recently, \citet{feain2011} have mapped the radio continuum structure of Cen~A at 1.4~GHz at high angular resolution and  \citet{osullivan} have produced a more detailed Faraday rotation analysis.

Low-frequency observations of the giant lobes of Cen~A have, until recently, suffered from poor angular resolution, e.g. the 408~MHz, 0\fdg85 resolution map of \citet{haslam} and the 45~MHz, 4\fdg6 resolution map of \citet{alvarez1997}. Low-frequency (below 100~MHz), low angular resolution studies of the Cen~A giant lobes have also been conducted by \citet{hamilton}, \citet{ellis}, \citet{shain} and \citet{sheridan}. A recent paper from the PAPER collaboration \citep{stefan} mapped Cen~A and the surrounding field at 148~MHz with 20-arcmin resolution and constructed a spectral-index map using the frequency variation within their 100~MHz observing band; they claim to find evidence of spectral flattening in turbulent regions of the lobes identified by \citet{feain2011}. Their spectral-index analysis is only qualitative, however, due to large uncertainties in their total flux measurements, resulting in physically unrealistic spectral index values of between 0 and -3 (see \citealt{stefan}, fig. 4).

 
\subsection{Previous spectral analyses}


The only spatially-resolved spectral-index maps of the giant lobes of Cen~A in the literature are by \citet{combi} and more recently by \citet{stefan}. The \citet{combi} map utilises single-dish data from Parkes at 408~MHz and 1.4~GHz and finds the spectral index to steepen at the edges of the northern lobe and to steepen with distance from the core in the southern lobe. However, when this spectral-index map is compared with the \citet{cooper1965} map of total intensity at 406~MHz, there is a correlation in the southern lobe between areas of steep spectral index and the estimated foreground component of emission from the `spur' feature of Galactic emission to the south-east \citep{cooper1965}. The spectral steepening in the northern lobe appears to correlate with regions of lower total intensity (and hence lower signal to noise against the Galactic foreground). \citet{stefan} attempt to map the spectral-index distribution of the giant lobes within the PAPER bandwidth and claim that the spectral index between 130 and 165~MHz varies as a function of distance from the core, steepening initially and then flattening toward the outer regions of the giant lobes. However, there appears to be a correlation between the total intensity and spectral-index maps in this case, indicating a possible systematic error in the calculation related to signal-to-noise. The error in the spectral-index maps is likely to be high in regions of low total intensity of the lobes where the bright Galactic foreground emission at low frequencies \citep{deoliveiracosta2008,bernardi} is a significant fraction of the emission, and particularly in the case of \citet{stefan} where the fractional frequency over which the spectral index is computed is very small. 

All other spectral-index calculations have been performed by taking ratios of flux densities summed over large regions, for example \citet{alvarez} and \citet{hardcastle2009}. This type of calculation is prone to error due to the inclusion of flux from foreground and background sources and constant offsets in maps due to missing short spacings. This is reflected in the large errors included in the spectral-index estimates of \citet{alvarez}. The most detailed spectral-index measurements of the giant lobes to date are from \citet{hardcastle2009}, who split the lobes into five large regions and use a combination of single-dish and 5-year WMAP data \citep{wmap2009} to constrain models of the spectral energy distribution (SED) and to make predictions as to whether the \emph{Fermi} Large Area Telescope (Fermi-LAT, \citealt{fermiLAT}) would detect gamma-ray emission from the lobes. Subsequent results from Fermi-LAT revealed strong gamma-ray detections in both the northern and southern lobes of Cen~A \citep{abdo,yang}.

\section{Observations and Data Reduction}

The MWA 32T was a scaled-down prototype version of the MWA, consisting of 32 antenna tiles, each tile consisting of 16 crossed-dipole antennas above a conducting mesh ground plane. The prototype was operated over the period September 2007 to September 2011 within the radio-quiet Murchison Radio Observatory in Western Australia at a latitude of -26\fdg7. Cen~A was observed with the MWA 32T over three consecutive nights from 2011 April 29 to 2011 May 1, using 80-s snapshot observations scheduled over a wide range of hour angles. Each snapshot observation used a different, fixed setting of the analog beamformers. The observations used in this paper consisted of a 30.72~MHz frequency band covering the frequency range of 108 to 139~MHz. The resulting dataset contained 80 snapshot observations, centred on 124~MHz, but only the lower 20~MHz was used in imaging for reasons that will be discussed below.  These snapshot observations were converted to UVFITS format and imported into {\sc casa} (Common Astronomy Software Applications) at a resolution of 1~s in time and 40~kHz in frequency. {\sc casa} was used for all further data reduction. {\sc aoflagger} \citep{offringa2010,offringa2012} was used to flag radio frequency interference (RFI) in each individual {\sc casa} measurement set. Additionally the central 40-kHz fine channel in each 1.24-MHz `coarse channel' and three fine channels at each coarse channel edge were flagged due to known rounding errors and aliasing, respectively. The measurement sets were then averaged to 4-s and 160-kHz resolution and concatenated into a single {\sc casa} measurement set.

We performed self-calibration on Cen~A itself, beginning with a simple, flat-spectrum point source model at the phase centre, while excluding baselines shorter than 100 $\lambda$. This was an effective calibration strategy, since the longer baselines do not resolve the bright central core of Cen~A, while excluding the shorter baselines removes the effect of the diffuse emission. A multi-frequency synthesis image of the entire 30.72~MHz bandwidth was made using a relatively shallow CLEAN in {\sc casa} in order to improve the sky model, and a single self-calibration iteration was then performed across this frequency band. The final image used a 20-MHz sub-band at a centre frequency of 118~MHz in order to have the widest (and simplest) primary beam shape. Use of only the lower 20~MHz of the bandwidth was found to reduce imaging artefacts around the centre of the source. It was found that natural weighting and a simple Cotton-Schwab CLEAN, without using the multiscale option, produced the best images. The W-projection algorithm \citep{wproj} was used to account for widefield effects.

\begin{figure*}
\centering 
\includegraphics[clip,trim=60 320 330 320,width=1.0\textwidth,angle=270]{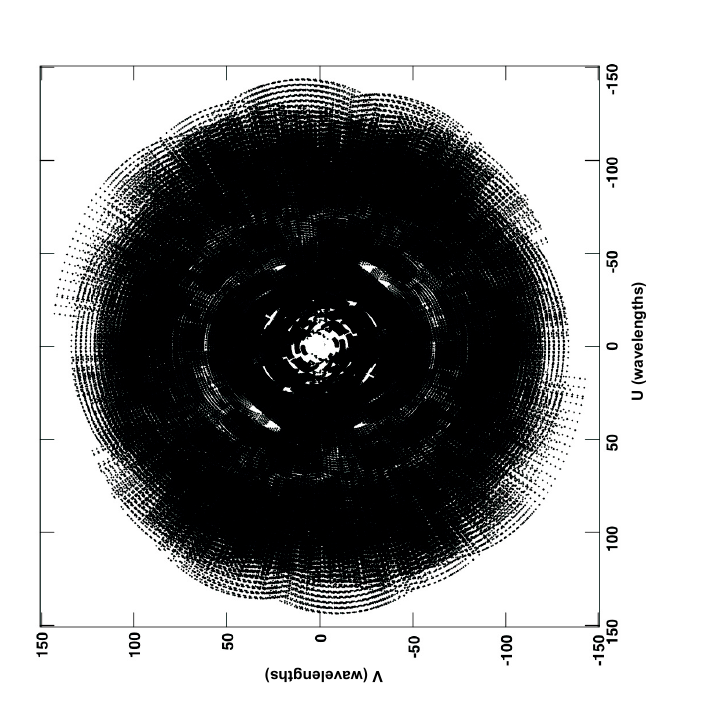}
\caption{Combined uv-coverage for all snapshots centred on 118~MHz with a bandwidth of 20 MHz, in units of wavelength. The plot does not include the three shortest baselines, which were excluded from calibration and imaging.}
\label{uv_118}
\end{figure*}

The imaging did not take into account the changing primary beam shape that varies as a function of frequency and time, making the beam and its polarisation properties different for each beamformer setting. This could be addressed using a Mosaic CLEAN if MWA primary beam shape images could be supplied to CLEAN. However, this functionality is currently not available and would require further development to be included in {\sc casa}. Our approach differs from that of \citet{williams}, who combined snapshots in the image domain, allowing them to assume a constant primary beam shape over each snapshot. The difference in approach is due to the scientific goal of imaging the large-scale structure associated with Cen~A and the foreground Milky Way with a very small number of short baselines. Earth rotation synthesis was required in order to increase the uv-coverage of the short spacings and to allow the CLEAN algorithm to reconstruct the large-scale structure. The combined uv-coverage for all snapshots at a centre frequency of 118~MHz, with a bandwidth of 20~MHz, is shown in Fig. \ref{uv_118}. 

\begin{figure*}
\centering 
\includegraphics[clip,trim=48 65 20 105,width=0.80\textwidth,angle=270]{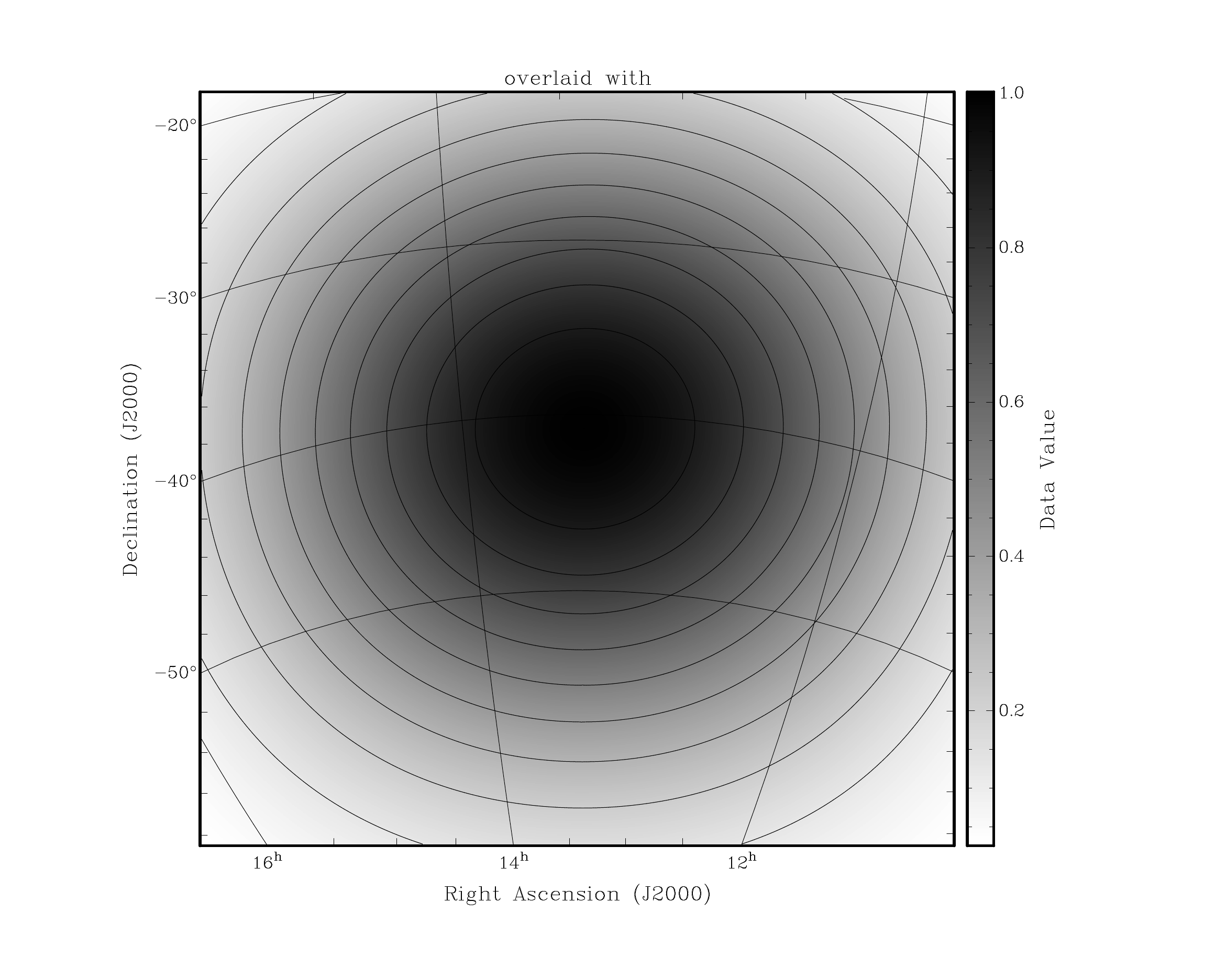}
\caption{Average tile primary beam from an analytic model. The centre frequency is 118~MHz. Contours are from 10\% to 90\% of the peak gain (normalised to 1), incrementing in steps of 10\%.}
\label{PB_118}
\end{figure*}

The MWA tile beam shape is different for both instrumental polarisations and varies as a function of frequency and time (due to different beamformer settings). An average `Stokes I' primary beam image was calculated by summing analytic primary beams calculated for each instrumental polarisation at each beamformer setting, using the same analytic primary beam model as \citet{williams}, and then taking the average of the two instrumental polarisation beamshapes. The average primary beam for the 118-MHz image is shown in Fig. \ref{PB_118}. The peak of the average primary beam in Fig. \ref{PB_118} is approximately $2\degr$ north of the core of Cen~A, due to the discrete nature of the beamformer settings used to electronically point the telescope tiles. The average primary beam correction was applied in image space to the final cleaned image. This approach assumes that each tile beam is the same, when in fact there will be some variation in the beam shapes between tiles. \citet{bernardi2013} have shown that the error introduced by the tile beam model used in MWA~32T observations at 189~MHz, also assuming the same beam shape for each tile, is at about the 2\% level.


Since self-calibration on Cen~A did not provide us with an overall flux scale, the flux scale was set using an ensemble of point sources within the full-width half-maximum (FWHM) of the primary beam. The Culgoora catalog \citep{culgoora,culgoora1995} is the only point source catalog covering this region of sky at the frequencies of interest. Eight unresolved Culgoora sources with measurements at both 80 and 160~MHz were identified within a $10\degr$ radius of Cen~A. A scaling factor was calculated for each of these sources by dividing its expected flux density (based on the Culgoora measurements at 80 and 160~MHz) by its measured flux density in the primary-beam-corrected MWA image. The details of the sources used to set the flux density scale in the MWA map at 118~MHz are given in Table \ref{fluxtable1}. The scaling factors are large because the flux density of the core of Cen~A was arbitrarily set to 1~Jy in the initial calibration. The standard deviation of the scaling factors listed in Table \ref{fluxtable1} is 319, which is 22\% of the average scaling factor of 1438. We therefore estimate the uncertainty in the flux scale of the MWA image of Cen~A to be $\pm22\%$. This is similar to the result of \citet{jacobs2013} who calculate a flux-scale uncertainty of 20\% for the MWA 32T over the full extent of its $\sim25\degr$ field of view and attribute most of this error to uncertainties in the primary beam model and to errors introduced by the CLEAN algorithm.


\begin{table}
\begin{minipage}{126mm}
\caption{Sources Used to Set Flux Density Scale}
\label{fluxtable1}
\begin{tabular}{@{}lcc} 
 \hline
 \shortstack{Source\\Name} & \shortstack{Expected Flux at\\118 MHz (Jy)} & \shortstack{Scaling\\Factor} \\
 \hline
PKS B1346-391 &  23.2 & 1946  \\
PKS B1246-410 &  31.6  & 1156 \\
PKS B1247-401 &  15.4  & 1185 \\
PKS B1340-373 &  6.6  & 1699  \\
PKS B1355-416 &  32.1 & 1237  \\
PKS B1243-412 &  13.7 & 1870 \\
PKS B1233-416 &  12.7 & 1153  \\
PKS B1407-425 &  13.4  & 1253 \\
\hline
  & \shortstack{Average \\Scaling Factor:} &  1438\\
  \hline
\end{tabular}
\end{minipage}
\end{table}

\section{Results and Analysis}

\begin{figure*}
\centering 
\includegraphics[clip,trim=20 0 40 15,width=0.85\textwidth,angle=270]{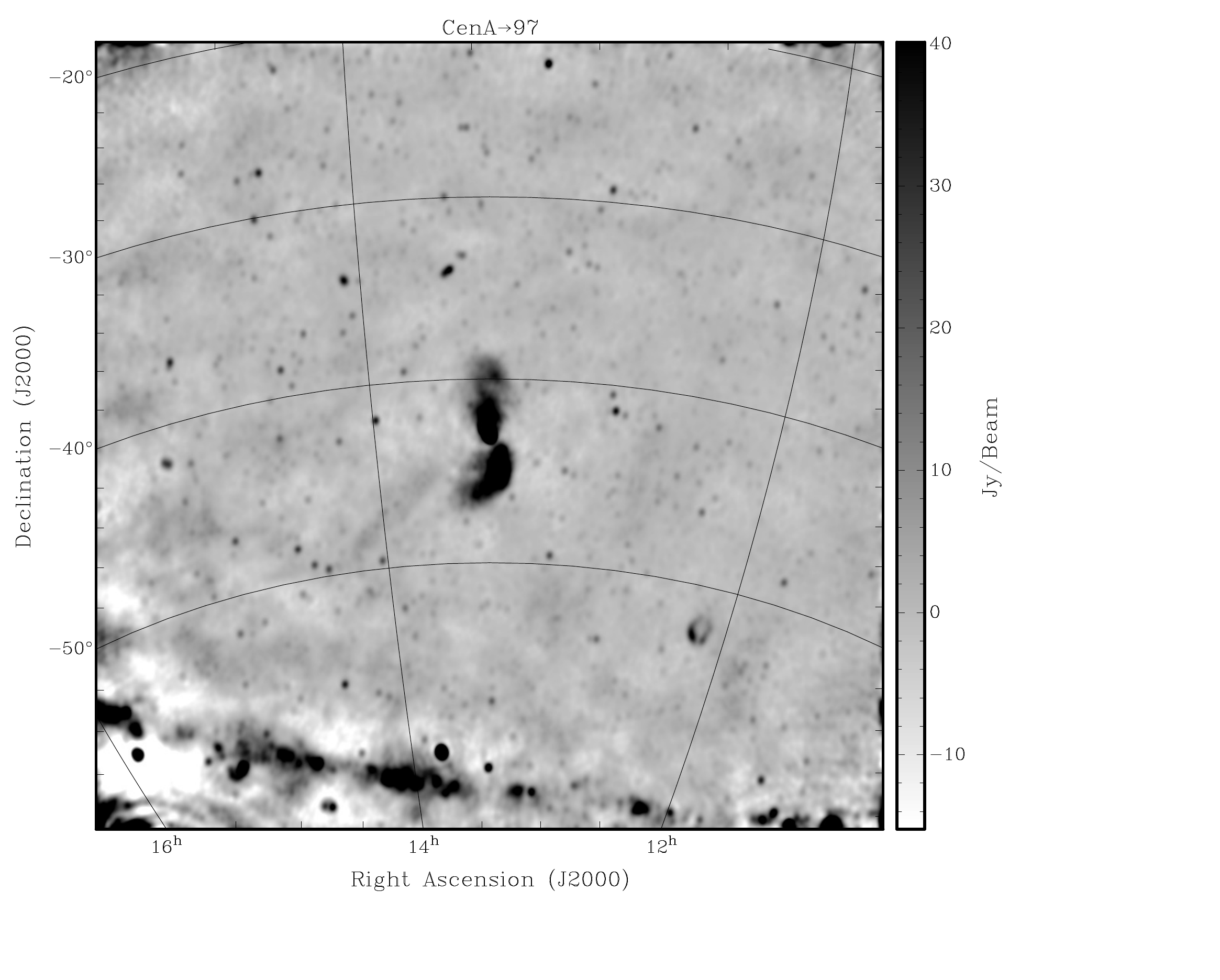}
\caption{Cen~A and surrounding field at~118 MHz with the Murchison Widefield Array 32-Tile prototype. The image is shown on a linear scale between $-15$ and $+40$ Jy/beam. It has an angular resolution of 25 arcmin and an rms noise of approximately 0.5 Jy/beam in `empty' regions of the map.}
\label{CenA118}
\end{figure*}

\begin{figure*}
\centering 
\includegraphics[clip,trim=20 80 52 60,width=0.65\textwidth,angle=90]{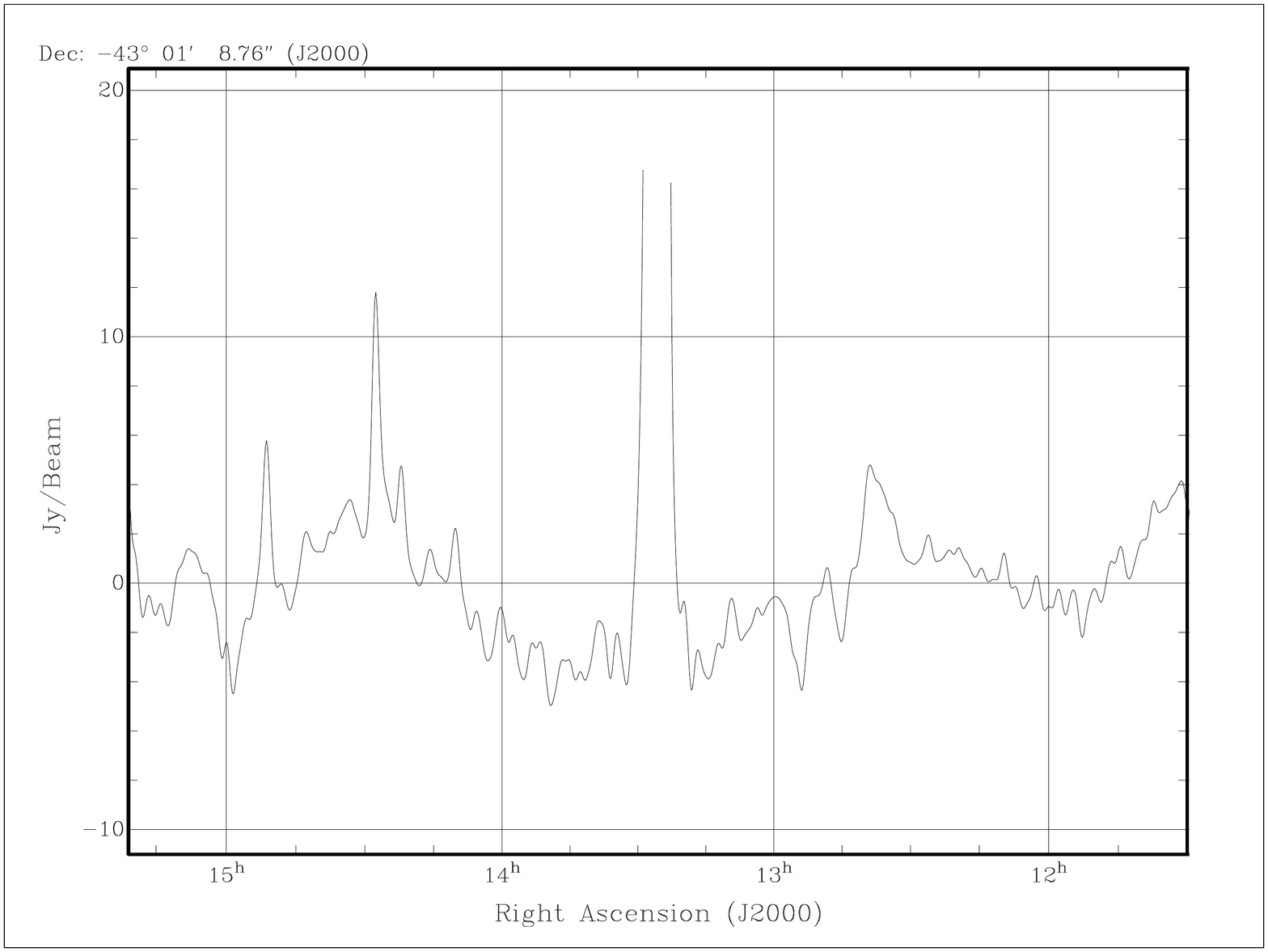}
\caption{Profile through the centre of the MWA 118~MHz image of Cen~A (see Fig.~\ref{CenA118}) along a constant declination of (J2000) $-43\degr01\arcmin9\arcsec$, showing the negative bowl surrounding the core of Cen~A in the image.}
\label{bowl}
\end{figure*}

Fig. \ref{CenA118} shows the MWA image of the Cen~A field at 118~MHz with the primary beam correction and flux scale applied. At this frequency, the spatial resolution is approximately 25~arcmin. The uv-range for this image was restricted to exclude the shortest three baselines of the MWA 32T that are problematic for calibration and imaging. This also effectively filters out features on scales larger than about $12\degr$, which would include much of the large-scale Galactic emission expected at these frequencies. Due to the similarities between PAPER and the MWA, many of the same imaging and calibration challenges described by \citet{stefan} have also been encountered in this work. We note that the missing short baselines in the MWA images result in a negative bowl around Cen~A, which affects the measurement of flux densities to some extent. Fig. \ref{bowl} shows a profile through the centre of the MWA image along a constant declination of (J2000) $-43\degr01\arcmin9\arcsec$. It appears from Fig. \ref{bowl} that the negative bowl, on the order of a few Jy/beam, is roughly constant within a $5\degr$ radius of the core of Cen A and hence has little effect on most of the subsequent analysis of Cen A's giant lobes. The total integrated flux density of the entire source in our 118~MHz image, measured by summing the flux in Cen~A above the 2.5~Jy/beam contour level, is 6620~$\pm$~1460~Jy. \citet{alvarez} show that the flux-density spectrum integrated over the full extent of Cen~A follows a power law with a spectral index of -0.70~$\pm$~0.01 from 10~MHz to 4.75~GHz and their fig.~2 predicts a flux density of approximately 7250~Jy. Our measured flux density for the entire source is consistent with the predicted value of \citet{alvarez}.


\begin{figure*}
$
\begin{array}{ccccc}\hspace{-2em}
\includegraphics[clip,trim=0 0 0 0,width=0.52\textwidth,angle=270]{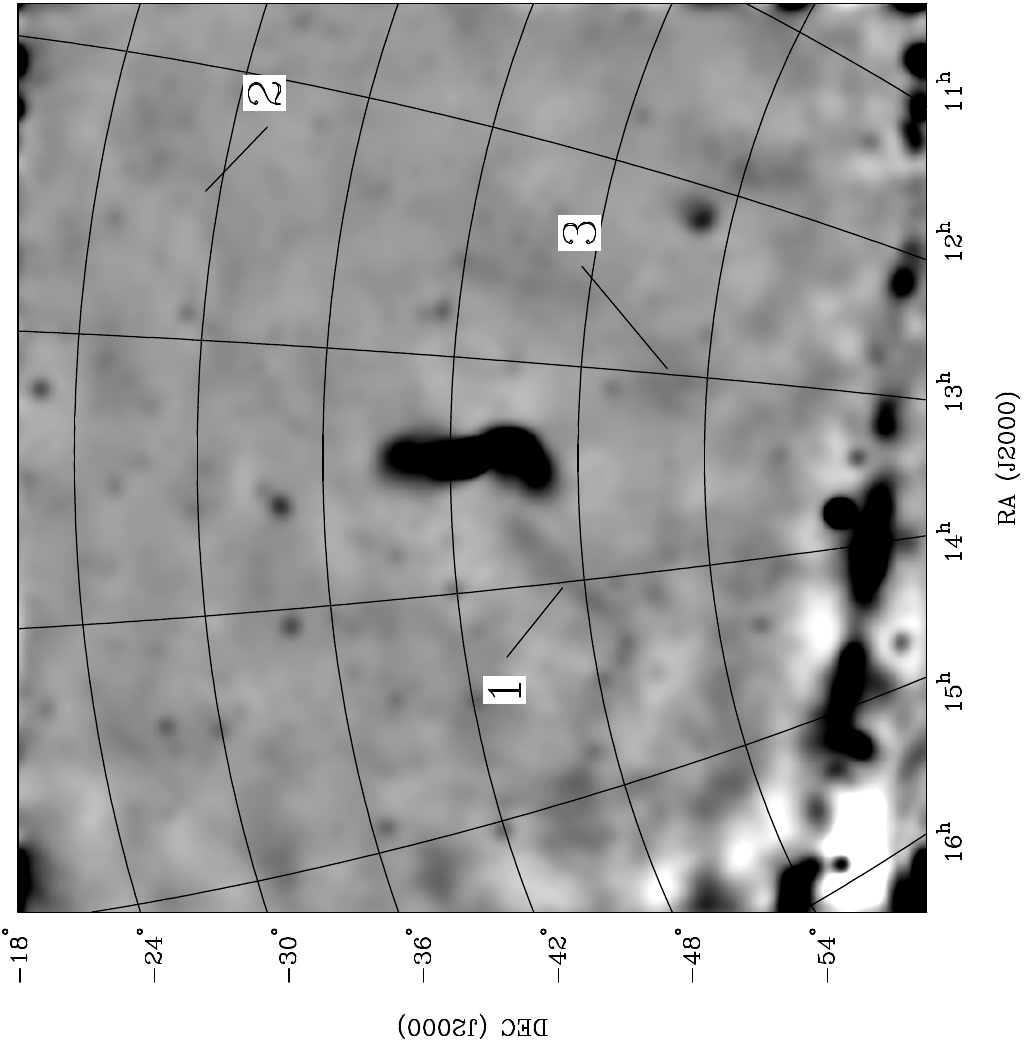} &
\includegraphics[clip,trim=0 35 0 0,width=0.52\textwidth,angle=270]{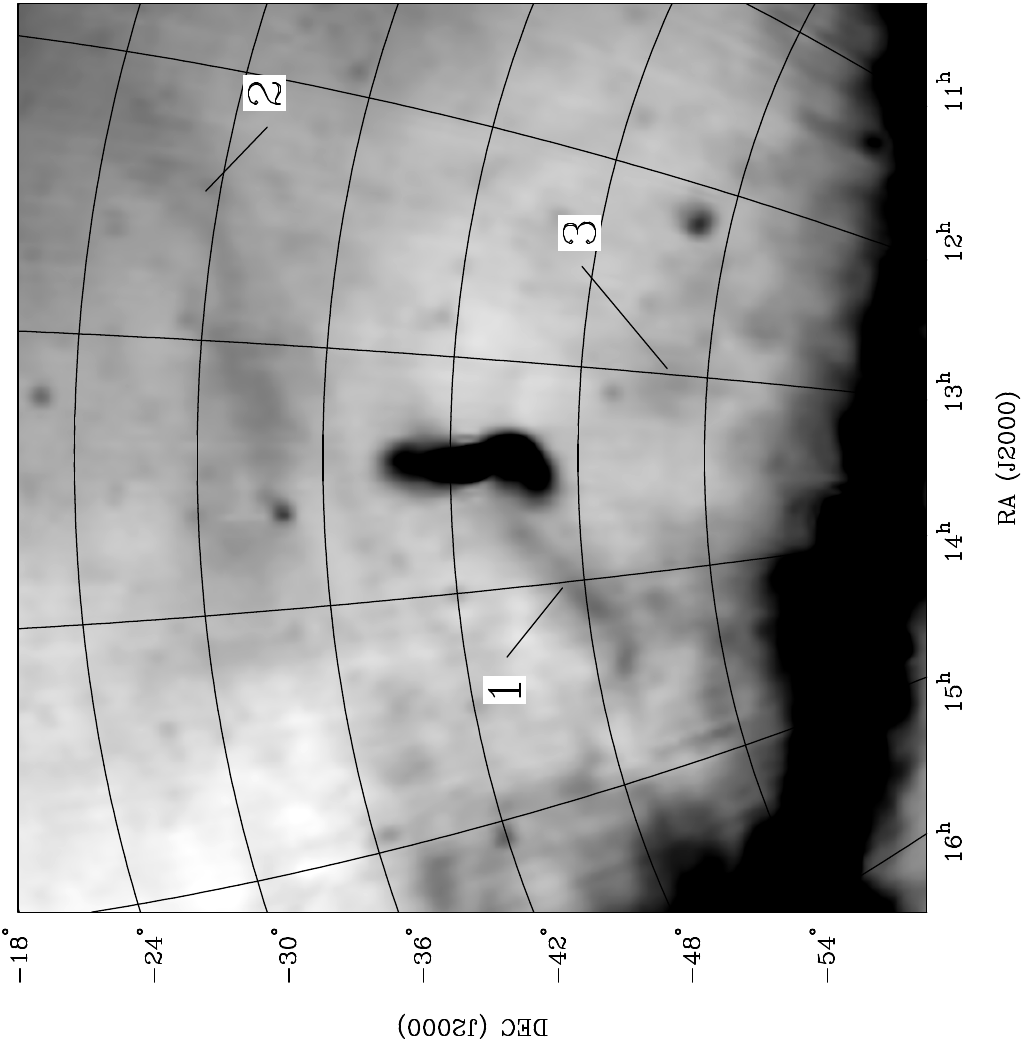}
\end{array}$
\caption{Comparison between the MWA 118~MHz image (left) and \citet{haslam} 408~MHz image (right), both smoothed to the same angular resolution of 0\fdg85. A linear brightness gradient has been subtracted from the \citet{haslam} map to make the smaller-scale structures clearer. The faint diffuse structure on scales less than $12\degr$ matches well by eye. Labels 1 to 3 indicate large-scale foreground features clearly present in both maps.}
\label{compare_has}
\end{figure*}

The peak flux density of Cen~A at 118~MHz is 1698~$\pm$~374~Jy/beam (where the Gaussian restoring beam is 0\fdg38~$\times$~0\fdg32 with a major axis position angle of $171\degr$) at RA (J2000) 13\textsuperscript{h}25\textsuperscript{m}28\textsuperscript{s}, Dec~(J2000)~$-43\degr01\arcmin9\arcsec$. This position is consistent with the known position of the centre of the radio source given by \citet{ICRF}. At  our angular resolution of 25~arcmin, the central resolution element contains the entirety of the inner lobes and the brightest part of the NML. The rms in a typical `empty' region of the image is approximately 0.5~Jy/beam, giving a dynamic range of at least 3000. At this frequency, the primary beam FWHM is approximately $31\degr$, and many faint point sources occupy the field of view. Apart from Cen~A, other distinctive features visible in Fig. \ref{CenA118} include the supernova remnant G296.5+10.0 in the south-west and the Galactic plane along the bottom of the image. The Galactic plane is not well reconstructed by the MWA~32T due to its large angular size, which is not sampled by the shortest baselines. Additionally, the Galactic plane is well outside the FWHM of the tile primary beam. Despite the relatively small number of short baselines, faint diffuse emission from the Milky Way on scales smaller than $12\degr$ can be seen across the map. To show that these large-scale structures in the surrounding field are real and not imaging artefacts, we show in Fig. \ref{compare_has} a comparison between the 408~MHz map of \citet{haslam} and the MWA 118~MHz map. The MWA image has been smoothed to the same resolution as the 408~MHz map (0\fdg85). The original \citet{haslam} data show a roughly linear decrease in average intensity from the south-east corner to the north-west (presumably due to Galactic emission). This large-scale gradient, undetectable to the shortest baselines of the MWA~32T, has been subtracted away from the \citet{haslam} map to make the smaller-scale structures clearer. In Fig. \ref{compare_has}, much of the diffuse structure between the two images matches well by eye, particularly the shape of the giant radio lobes of Cen~A themselves and the `spur' feature to the south-east (label 1). Labels 2 and 3 in Fig. \ref{compare_has} show two other distinctive large-scale foreground features that appear in both maps. The qualitative comparison between the \citet{haslam} and MWA maps confirms that the features on the angular scales associated with the Cen~A giant lobes are being reconstructed properly in the MWA image. We do not pursue a spectral comparison with the Haslam map as the angular resolution is too coarse and the frequency difference between the two maps is too small for our purposes. Instead, we use data at 1.4~GHz for our spectral analysis.


\begin{figure*}
\centering 
\includegraphics[clip,trim=63 160 50 160,width=1.0\textwidth,angle=270]{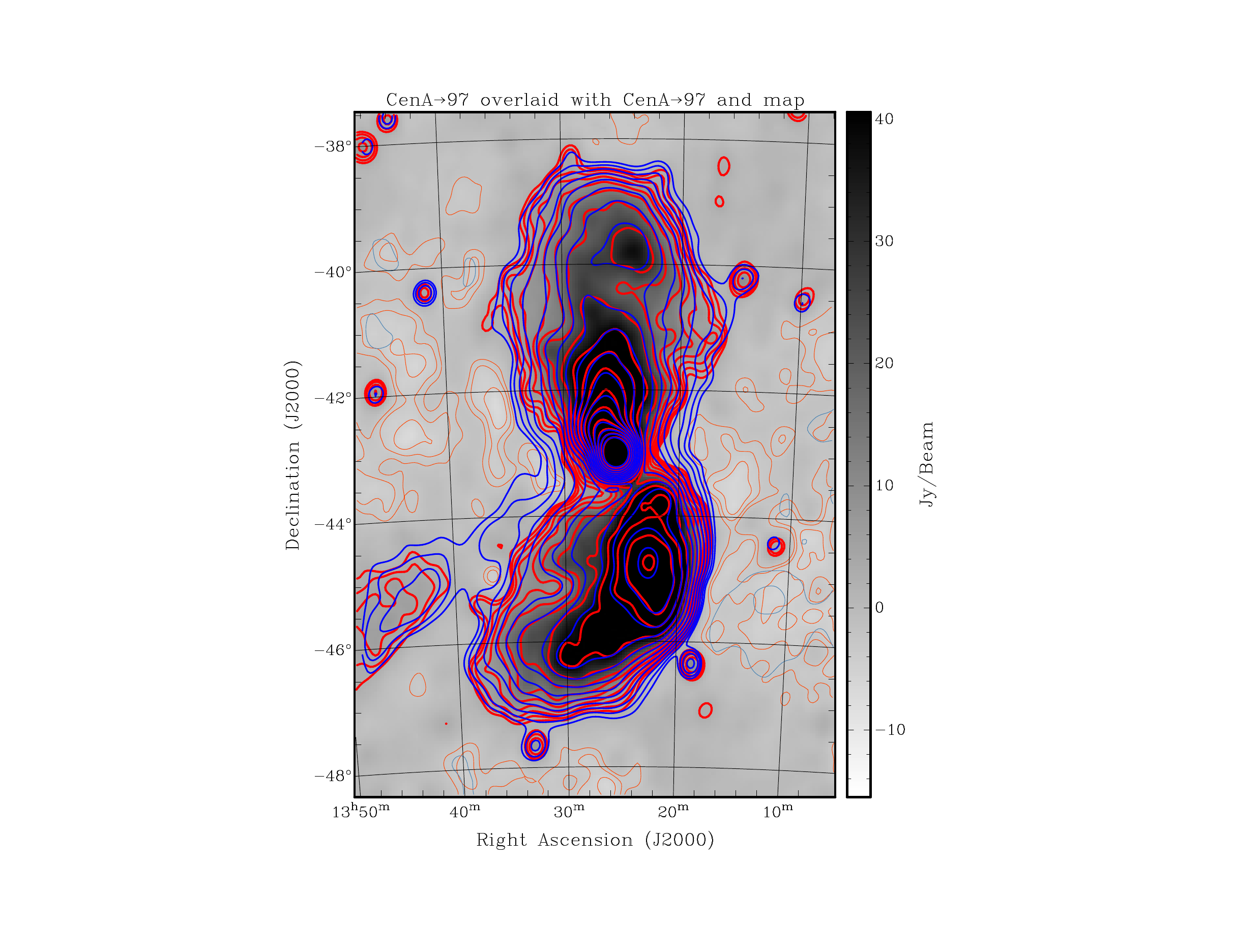}
\caption{Cen~A at 118~MHz (greyscale and red contours) overlaid with Parkes 1.4~GHz contours in blue, smoothed to the same angular resolution of 25~arcmin. Positive contours are bold and start at 2.5~Jy/beam for 118~MHz and 0.50~Jy/beam for 1.4~GHz, incrementing in a geometric progression of $\sqrt{2}$. Negative contours (thinner and lighter shade) start at -2.5~Jy/beam for 118~MHz and -0.50~Jy/beam for 1.4~GHz and decrement in a geometric progression of $\sqrt{2}$.}
\label{CenAcompare}
\end{figure*}

Fig. \ref{CenAcompare} shows the MWA image at 118~MHz (in both greyscale and contours) overlaid with the Parkes 1.4~GHz Stokes I contours from \citet{osullivan}. The 1.4~GHz map has been smoothed to the same resolution as the MWA map. The structure of Cen~A at 118~MHz matches very closely with the structure seen at 1.4~GHz. However there are a few areas where the morphologies appear to be slightly different. Of particular interest is the northern part of the southern giant lobe at approximately RA (J2000) 13\textsuperscript{h}23\textsuperscript{m}, Dec (J2000) $-43\degr45\arcmin$, where there appears to be additional structure at 118~MHz that is not present in the 1.4~GHz image. This is discussed further in combination with the spectral tomography results in Section 4.1.1 below.

\subsection{Spectral index of the giant lobes}

Calculation of the spectral index only within the 30~MHz MWA band is unlikely to produce accurate values since the fractional frequency coverage is too small to accurately measure changes in flux density. Instead, we compare data at 1.4~GHz \citep{osullivan} to our 118~MHz image in order to compute spectral indices. This larger frequency ratio makes the spectral-index calculations more robust to errors in both maps. The assumption of a single spectral index over such a large frequency range is simplistic as there could be curvature in the spectrum. However, between 118~MHz and 1.4~GHz the spectral curvature is expected to be small, as predicted by the inverse-Compton modelling of \citet{yang}; for example see their fig. 8. 

We first measure the spectral indices of the giant lobes over the same regions as \citet{hardcastle2009} using T-T plots. This technique accounts for the possibility of constant offsets between maps due to missing short spacings or large-scale foreground contamination. T-T plots were initially used in the analysis of astronomical observations to account for a constant but unknown component of antenna temperature being contributed by the ground in observations of the spectrum of Galactic radio emission \citep{turtle}. The technique takes the brightness temperature of many points within a region of the sky (assumed to have the same spectral index) at two different frequencies and plots them against each other. If there is a constant offset between the brightness temperature values within a region due to emission from the ground (or any other reason) then this will be obvious in a T-T plot; the points in a T-T plot between frequencies $\nu_1$ and $\nu_2$ will fall along a line having a slope equal to $(\nu_1/\nu_2)^{\alpha}$, and with a non-zero y-intercept resulting from the constant offset between the two regions. At 118~MHz, the brightness temperature of the Galactic foreground is two orders of magnitude higher than at 1.4~GHz and if significant on the spatial scales of our chosen regions, it would show up as a constant offset in T-T plots between 118~MHz and 1.4~GHz. The effects of missing flux due to missing short baselines in the interferometer data would also cause a constant offset in the T-T plots.

\begin{figure*}
\centering
\includegraphics[clip,trim=30 100 0 120,width=0.80\textwidth,angle=270]{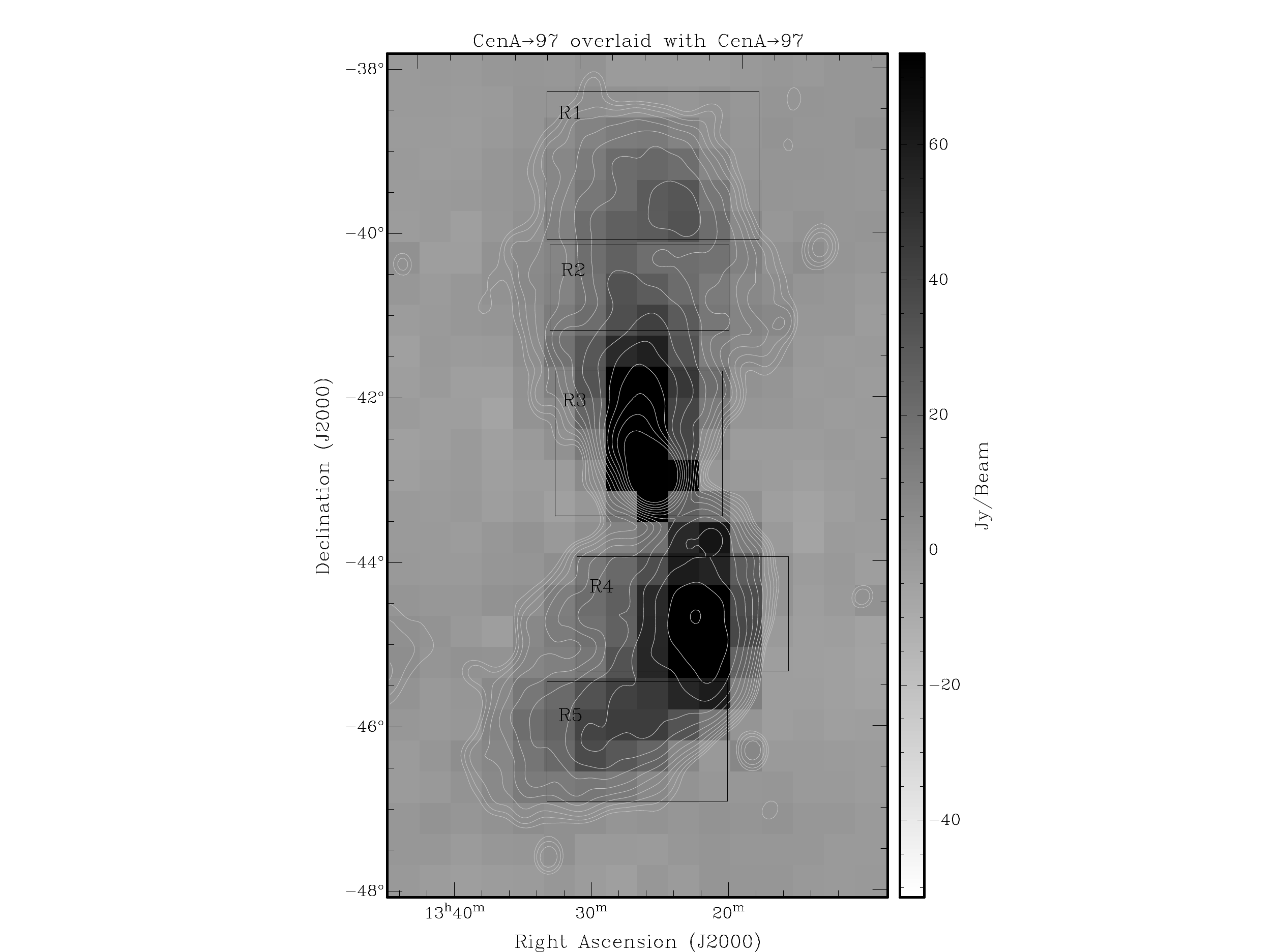}
\caption{Regions from \citet{hardcastle2009} used for constructing T-T plots, overlaid on the MWA 118~MHz image resampled to 1 pixel per beam. Contours are of the full resolution 118~MHz image, incrementing in a geometric progression of $\sqrt2$ from 2.5 Jy/beam. }
\label{regions}
\end{figure*}

We constructed T-T plots for the five regions shown in Fig. \ref{regions}, chosen to match with the regions defined by \citet{hardcastle2009}. Our region 3 differs from \citet{hardcastle2009} in that we use a rectangle, rather than a circle, to define this region. The 1.4~GHz images were smoothed to the same 25-arcmin resolution as the MWA map (Fig. \ref{CenA118}) and both images were resampled to make the pixel size as large as the synthesised beam. This produced independent measurements of the flux density at each (equally sized) pixel in both images, allowing us to perform a chi-squared analysis. For each region, the flux densities at the low and the high frequencies at each pixel were plotted against each other. The error bars for each point were estimated by selecting off-source regions to the east and west of each of the regions in Fig. \ref{regions} and measuring the rms noise in each. These off-source regions were large enough to ensure that the rms measurement included both variations in Galactic foreground emission and the contribution of background point sources. The reduced chi-squared ($\chi_{red}^{2}$) was also computed for each T-T plot and used as an indicator of the validity of the assumption of a single spectral index over the spatial extent of a given region. A line was fit to the data using orthogonal-distance regression \citep{ODR} and the spectral index for each region,  $\alpha_{R}$, was calculated using:
\begin{equation} 
\alpha_{R}=\frac{\log{(\Delta)}}{\log{(\nu_{low}/\nu_{high})}},
\label{alpha2}
\end{equation}
where $\Delta$ is the slope of the line of best fit for the T-T plot of region $R$ and $\nu_{low}$ and $\nu_{high}$ are the low and high frequencies, respectively.

\begin{figure*}
\centering 
\includegraphics[clip,trim=20 150 20 150,width=1.0\textwidth,angle=270]{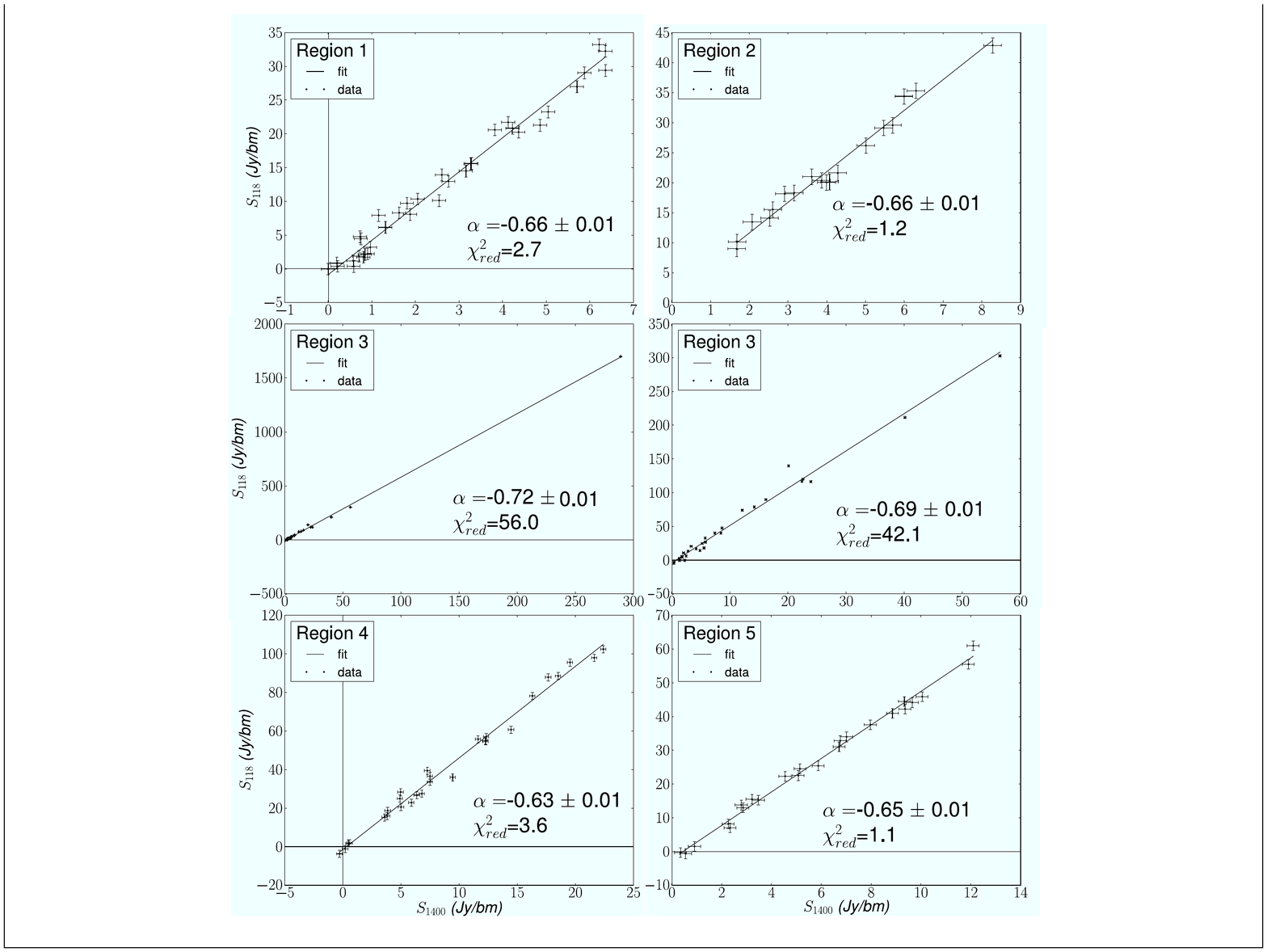}
\caption{T-T plots for the five regions of the giant lobes of Cen~A and the computed spectral indices between 118~MHz and 1400~MHz, including the  $\chi_{red}^{2}$ values for each region as defined in Fig. \ref{regions}. The centre row contains two plots for region 3; on the left all pixels have been included, and on the right the bright central pixel has been excluded to remove the influence of the inner lobes. The uncertainties in the spectral indices displayed are derived from the uncertainty in the slope of the line of best fit and represent random errors only. See text for the inclusion of systematic errors due to global flux scale uncertainties.}
\label{tt2}
\end{figure*}

Fig. \ref{tt2} shows T-T plots for the five regions of the giant lobes and the computed values of $\alpha$ and  $\chi_{red}^{2}$ for each. The centre row of Fig. \ref{tt2} contains two plots for region~3; on the left all pixels have been included, and on the right the bright central pixel has been excluded to remove the influence of the inner lobes, since we are primarily concerned with the spectral indices of the giant outer lobes in this paper. The $\chi_{red}^{2}$ values for regions 2 and 5, corresponding to the inner part of the northern outer lobe and the outer part of the southern outer lobe, respectively, are close to unity, indicating that the spectral indices of these regions show little spatial variation. The T-T plots for region~3 have $\chi_{red}^{2}$ values of 56.0, if the central pixel is included, and 42.1, if the central pixel is excluded. These $\chi_{red}^{2}$ values are an order of magnitude larger than for the other regions, indicating significant spatial variation of the spectral index. This is probably due to the complex structure of the region that contains the NML. Regions 1 and 4 were also found to have high $\chi_{red}^{2}$ values of 2.7 and 3.6, respectively. The variation in spectral index in these regions will be investigated in detail using spectral tomography in Section~4.1.1. 

The spectral indices derived from the T-T plots can be considered as averages over each region and so can be compared to the values computed using summed flux densities in previous studies. The main source of error in the spectral indices is due to the 22\% uncertainty of the global flux scale in the MWA 118~MHz image. This error does not affect differences in the spectral indices computed between regions. We therefore quote separately the systematic errors due to global flux scale uncertainties and the statistical errors due to random errors in the images. The error in the flux scale of the 1.4~GHz data is approximately 2\% \citep{reynolds1994,baars} and is negligible compared to the uncertainty at 118~MHz. We find a spectral index in regions 1 and 2 (northern outer lobe) of $\alpha_{1,2} = -0.66 \pm 0.01{\rm (stat)}^{+0.10_{\rm sys}}_{-0.08_{\rm sys}}$, and in regions 4 and 5 (southern outer lobe) of $\alpha_{4} = -0.63 \pm 0.01{\rm (stat)}^{+0.10_{\rm sys}}_{-0.08_{\rm sys}}$ and $\alpha_{5} = -0.65 \pm 0.01{\rm (stat)}^{+0.10_{\rm sys}}_{-0.08_{\rm sys}}$, respectively. The calculated spectral-index values are consistent with \citet{alvarez}, who found the spectral indices of the northern and southern lobes between 408~MHz and 1.4~GHz to be $-0.75\pm0.08$ and $-0.70\pm0.08$, respectively. Our values are also consistent with the \citet{hardcastle2009} spectral indices between 408~MHz and 1.4~GHz (which we calculated from the flux density values in their Table 1) of $\alpha_{1}=-0.60\pm{0.05}$ and $\alpha_{2}=-0.65\pm{0.07}$ in the northern lobe and $\alpha_{4}=-0.47\pm0.06$ and $\alpha_{5}=-0.62^{+0.12}_{-0.11}$ in the southern lobe. 




As well as being averages across a given angular region, the spectral indices computed using T-T plots are averages along our line of sight. This presents a problem if there are overlapping regions with different spectral indices. In a scenario where the lobes were formed by multiple, separate outbursts from the central engine \citep{morganti,saxton}, there could be overlapping structures of synchrotron-emitting particles produced during different epochs of activity, as discovered in another relatively close-by radio galaxy Hercules A \citep{HerA}. In this case, the restarted jets forming the more recent lobe structures have a faster advance speed than the previous jets since they are travelling through an underdense region that has been cleared of material in the previous epoch of activity \citep{clarke1991}. As the new lobes catch up to the older lobes, this results in particles with different spectral ages (and presumably different spectral indices) overlapping along our line of sight. To investigate this possibility further we apply the technique of spectral tomography in Section~4.1.1.

\subsubsection{Spectral Tomography}

The simple two-point spectral index generated by dividing images at different frequencies is unable to separate structures with different spectral indices when they are aligned along our line of sight, instead giving a weighted average of the two. Spectral tomography has been used to identify overlapping spectral structures within radio galaxies \citep{katz-stone,HerA} and has also been applied to Galactic sources \citep{delaney,gaenslerwallace}. Here we apply the technique to the giant lobes of Cen~A to search for regions with distinctly different spectral indices, which may overlap along our line of sight and therefore may be difficult to detect using conventional methods.

Since the spectral index of each component is not known, a cube of images is generated, each with a different spectral component subtracted away. The cube of images, $M_{tom}(\alpha_{t})$, is given mathematically by:
\begin{equation} 
M_{tom}(\alpha_{t})\equiv M_{low}-\left(\frac{\nu_{low}}{\nu_{hi}}\right)^{\alpha_{t}}M_{hi},
\label{tomeqn}
\end{equation}
where  $M_{low}$ and  $M_{hi}$ are the low and high frequency maps, $\nu_{low}$ and $\nu_{hi}$ are the low and high frequencies, respectively, and ${\alpha_{t}}$ is the trial spectral index for that image in the cube. 

When the trial spectral index is correct, the component is subtracted perfectly and the region of the image appears to match with the background. Whenever the trial index is incorrect, the component is either under-subtracted or over-subtracted, making it appear lighter or darker than the background. The uncertainty in the spectral index is determined by the range of trial values in which a feature on the map is neither under-subtracted nor over-subtracted. The uncertainties quoted in this section are therefore due to random errors only and do not include the systematic error introduced by the uncertainty in the global flux scale. We generated a spectral tomography cube with trial spectral indices from -0.01 to -1.50, decrementing in steps of 0.01. When the tomography cube is viewed as a movie, it is clear that there are variations in spectral index with position across both the northern and southern outer lobes of Cen~A. We show a subset of the images from the cube in Figs \ref{tom1north} and \ref{tom2south} below.

First we examine the northern lobe, and in particular region 1 where the $\chi_{red}^{2}$ of the T-T plot indicated that a power law with a single spectral index for the whole region was not a good model (see Fig. \ref{tt2}). A series of 6 slices of the tomography cube is shown in Fig.~\ref{tom1north}, with trial spectral indices from $\alpha_t=-0.45$ to $\alpha_t=-0.70$, decrementing in steps of 0.05. The spectral tomography reveals that the spatial distribution of spectral-index values is complex, and although the average value of around -0.65 (lower, centre panel) agrees well with the T-T plots, it is clear that there are a range of spectral index values distributed across region 1 and across the rest of the northern lobe. In particular, the spectral index flattens significantly at a declination of around $-39\fdg5$ (label A) where the position angle of the northern lobe changes to form the outer northern-hook structure. From the tomography cube, we estimate that the spectral index of this flattened region is $-0.55\pm0.05$ and steepens to $-0.70\pm0.05$ at the end of the northern hook (label B). An interesting result is that a similar positional change in spectral index also occurs in the region closer to the core. Immediately south of declination $-42\degr$ there is a relatively flat region of spectral index $-0.67\pm0.01$ (label D) which abruptly steepens to $-0.72\pm0.01$ north of declination $-42\degr$ (label C). These features are most clearly seen where they are slightly over-subtracted or under-subtracted, hence the reason for showing slices at  $\alpha_t=-0.65$ and $\alpha_t=-0.70$ in Fig.~\ref{tom1north}, rather than $\alpha_t=-0.67$ and $\alpha_t=-0.72$. There is also a distinctly steeper region at 13\textsuperscript{h}31\textsuperscript{m}18\textsuperscript{s}, $-41\degr19\arcmin54\arcsec$, however this is not a feature of the lobes of Cen~A. It is due to the background point source PKS~1328-410 \citep{pks1328-410}, which can also be seen clearly in the high resolution map of \citet{feain2011}.

A subset of the spectral tomography cube zoomed in on the southern outer lobe is shown in Fig.~\ref{tom2south}. Although the large-scale morphology is quite different from the northern lobe, we do see similar changes in spectral index with position in the south. This is most prevalent in the northern region of the southern lobe (corresponding to region 4 in the T-T plots in Fig.~\ref{tt2}). Of interest is the distinctly steeper region at approximately Dec (J2000)~$-43\degr45\arcmin$ (label E) with a spectral index of $-0.69\pm0.02$ surrounded by regions of spectral index $-0.57\pm0.02$ immediately to the north and south (label F). This is best seen in the lower-left panel ($\alpha_t=-0.59$) of Fig.~\ref{tom2south}, where the steeper feature is under subtracted, appearing darker than the flatter-spectrum region surrounding it. The feature labelled E in Fig.~\ref{tom2south} appears to be present at 118~MHz but not at 1.4~GHz. To exclude the possibility that this could be due to a bright, steep-spectrum background point source, we have plotted in Fig.~\ref{SML}, the MWA and Parkes contours over the high resolution ATCA+Parkes map of \citet{feain2011}. Fig.~\ref{SML} shows that the additional peak at RA (J2000) 13\textsuperscript{h}23\textsuperscript{m}, Dec (J2000) $-43\degr51\arcmin$ in the 118~MHz map does not coincide with any background sources visible at 1.4~GHz. The bright extended source MRC~1318-434B \citep{schilizzi,pks1328-410} is clearly visible in the high resolution greyscale image and is largely responsible for the peak in the contours of the smoothed 118~MHz and 1.4 GHz images at RA (J2000) 13\textsuperscript{h}21\textsuperscript{m}, Dec (J2000) $-43\degr48\arcmin$, which is approximately 30~arcmin away from the additional peak in the 118~MHz image. The positional registration of the two images is accurate to within 1~arcmin. Two other steep spectrum features that are prominent in Fig.~\ref{tom2south} at RA (J2000) 13\textsuperscript{h}29\textsuperscript{m}38\textsuperscript{s}, Dec (J2000) $-44\degr44\arcmin14\arcsec$ and RA (J2000) 13\textsuperscript{h}33\textsuperscript{m}57\textsuperscript{s}, Dec (J2000) $-46\degr41\arcmin32\arcsec$ are due to the background point sources SUMSS J132937-444413 \citep{sumss} and PMNM 133051.5-462614 \citep{pks1328-410}, respectively. Again, these bright background sources are clearly visible in the high resolution map of \citet{feain2011}.

\begin{figure*}
\centering 
\includegraphics[clip,trim=20 100 20 50,width=0.92\textwidth,angle=270]{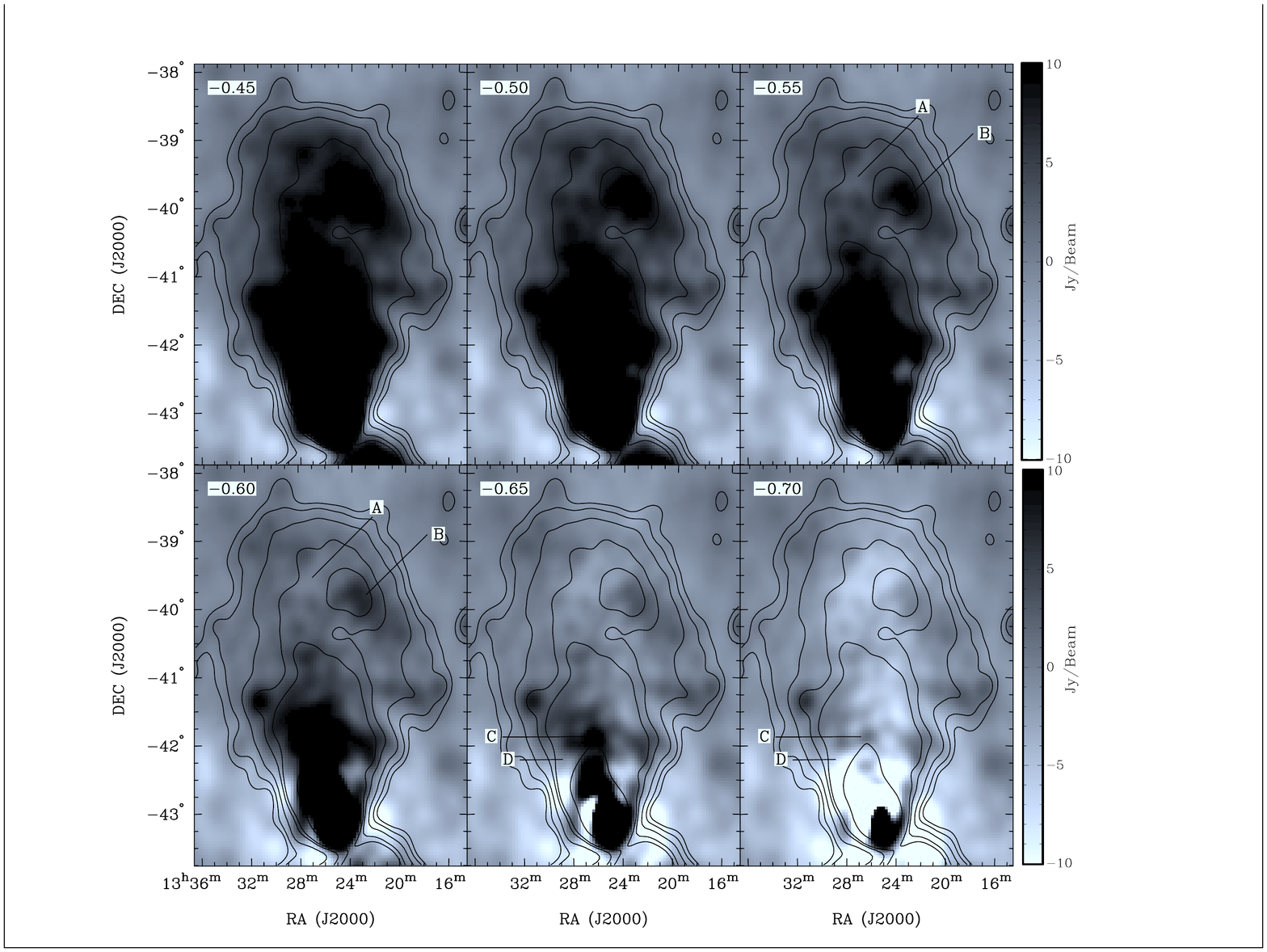}
\caption{A subset of the spectral tomography images of the northern lobe between 118~MHz and 1.4~GHz with trial spectral indices as indicated in the top-left corner of each sub-image. Greyscale is from -10 to 10 Jy/beam. Overlaid are contours from the 118~MHz image at 2.5, 5, 10, 20, 28.3 and 113 Jy/beam. The feature labelled A has a significantly flatter spectral index than that labelled B. Similarly, there is a change in spectral index between the flatter region labelled D and the steeper feature labelled C.}
\label{tom1north}
\end{figure*}

\begin{figure*}
\centering 
\includegraphics[clip,trim=20 150 20 85,width=0.92\textwidth,angle=90]{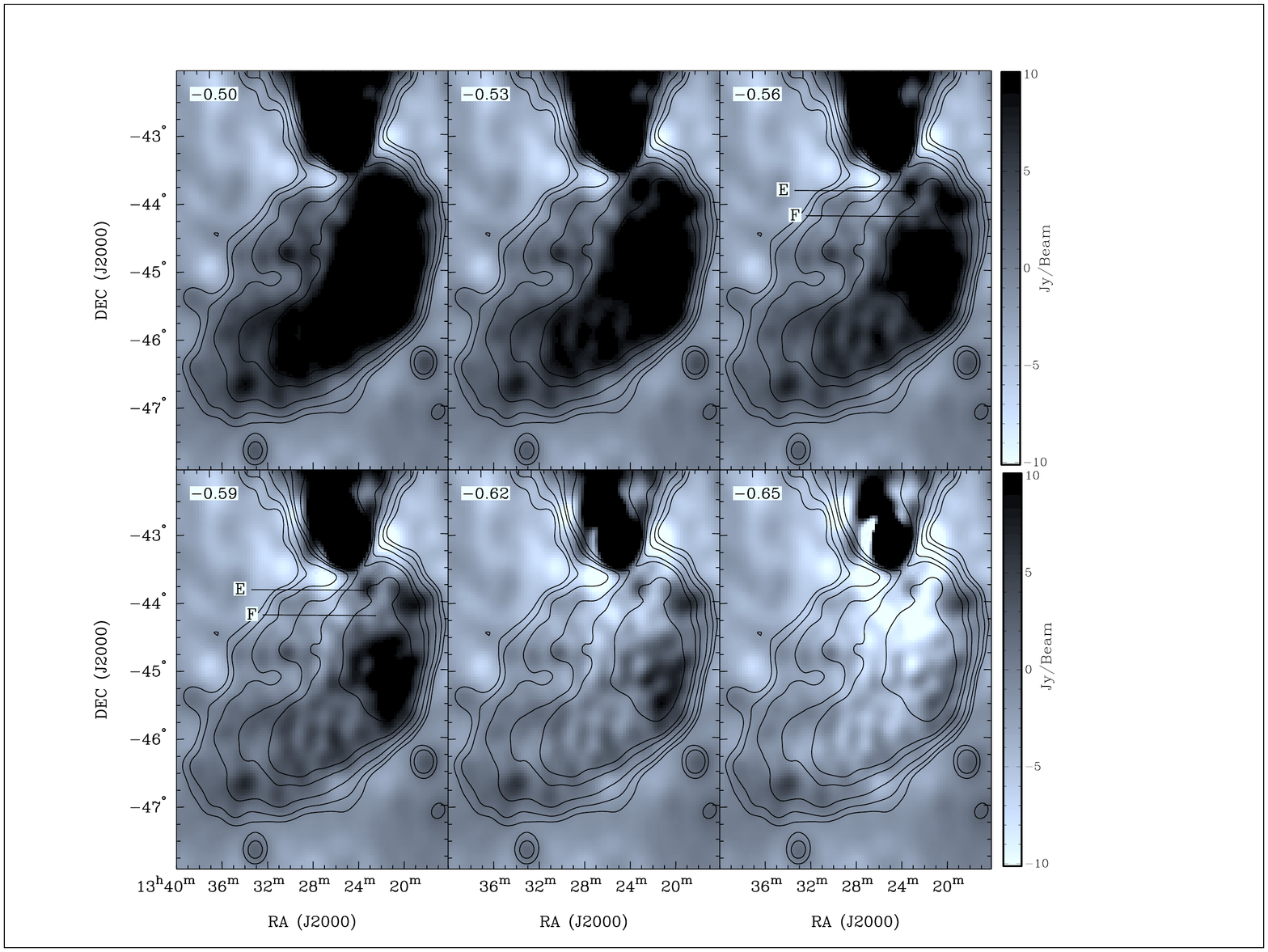}
\caption{As for Fig.~\ref{tom1north}, but for the southern lobe with contours from the 118~MHz image at 2.5, 5, 10, 20, 28.3 and 56.5 Jy/beam. The feature labelled E has a significantly steeper spectral index than that labelled F.}
\label{tom2south}
\end{figure*}

\begin{figure*}
\centering 
\includegraphics[clip,trim=10 0 0 0,width=0.75\textwidth,angle=270]{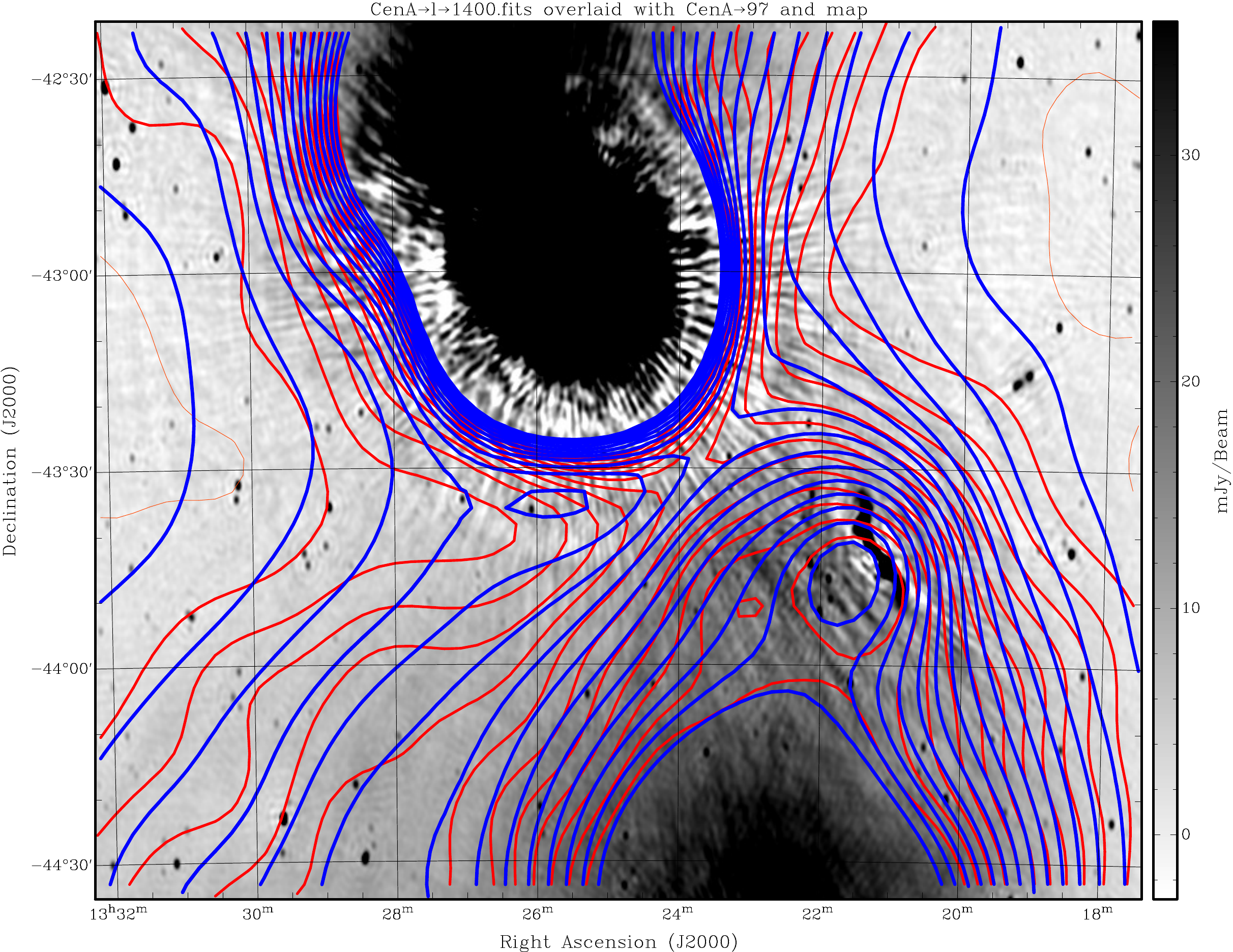}
\caption{High resolution 1.4~GHz Parkes+ATCA map \citep{feain2011} in greyscale, overlaid with the 1.4~GHz Parkes contours \citep{osullivan} in blue and the MWA 118~MHz contours in red. Positive 118~MHz MWA contours are from 3~Jy/beam incrementing in steps of 5 Jy/beam and 1.4~GHz Parkes contours are from 1~Jy/beam incrementing in steps of 1~Jy/beam. There is one negative contour visible, which is the $-3$ Jy/beam level of the 118~MHz map, shown as a thinner, lighter red line. There is a peak at 13\textsuperscript{h}23\textsuperscript{m}, $-43\degr51\arcmin$ in the 118~MHz map that is not visible in the 1.4~GHz images and does not coincide with any background sources visible at 1.4~GHz. The bright background source MRC~1318-434B, at 13\textsuperscript{h}21\textsuperscript{m}18\textsuperscript{s}, $-43\degr41\arcmin15\arcsec$, extends approximately 20~arcmin and is coincidentally aligned with the position angle of the inner lobes of Cen~A.}
\label{SML}
\end{figure*}

\section{Discussion}

\subsection{Morphology, spectral index and clues about lobe formation history}

The morphology of Cen~A at 118~MHz with 25-arcmin resolution matches closely with the single dish Parkes observations at 1.4~GHz, as shown in Fig.~\ref{CenAcompare}. The 15~arcmin gap between the northern and southern lobes reported by \citet{feain2011} can also be seen in the 118~MHz image, although in the latter data the emission does not quite disappear down to the noise level, since the synthesised beam of 25~arcmin is too large to fully resolve this feature. The difference in the large-scale morphologies between the northern and southern lobes is quite distinct, with the northern lobe characterised by an abrupt change in position angle producing the northern hook feature and the southern lobe having a much smoother variation in orientation. These large-scale features mirror the morphology of the inner lobes as imaged at high spatial resolution with the Very Large Array \citep{schreier,vla}. In the north, a sudden change in position angle of the radio emission is also seen at an intermediate scale in the Northern Middle Lobe (NML), as imaged in detail by \citet{morganti} (see e.g., their fig. 4). An explanation for the existence of the intermediate scale NML and the fact that it mirrors the shape of the inner lobes was proposed by \citet{morganti}, who suggested that the central engine has undergone multiple outbursts of activity. In this scenario, the inner lobes are due to the most recent outburst and the middle and outer lobes are relics of past activity. They suggest that the distinctive shape of the NML and its associated large scale jet could be due to a similar type of interaction with the surrounding environment that produces the class of radio galaxies known as wide-angled tailed (WAT) radio galaxies. \citet{wats} used numerical simulations of jets disrupted by shocks in the ambient medium to reproduce the structure of WATs and the inner northern jet of Cen~A. If the inner lobes are the result of the current outburst of activity from the core and the NML is the result of a previous outburst, then the outer northern lobe with its hook-like structure seen at frequencies below 5~GHz is probably the result of similar previous outbursts of AGN activity. In order to build up such large, energetic structures as the giant outer lobes, many outbursts similar to the current inner lobe activity are likely to have been required.


Clear evidence of multiple, separate outbursts of activity in radio galaxies is seen in the class of radio galaxies known as `double-double' radio galaxies (see e.g., \citealt{schoenmakers2000,jamrozy}). However, these are characterised by edge-brightened lobes like those seen in FRII galaxies. \citet{jamrozy} single out several other radio galaxies that show evidence of multiple outbursts of activity including 3CR310 \citep{leahy}, the WAT galaxy in the cluster A~2372 \citep{gia}, Hercules A (Her~A, \citealt{HerA}) and also Cen~A. Spectral-index maps of the WAT galaxy in A~2372 and Her~A show similar features to our spectral-tomography results, where the spectral index changes with distance from the core, showing alternating regions of flatter and steeper spectral index indicating different spectral ages across the lobes. In our spectral-tomography results for the northern lobe of Cen~A (Fig.~\ref{tom1north}), we show that the spectral index steepens along the extended structure associated with NML from point D to point C, indicating an older population of particles further from the core. The spectrum then flattens again at point A, which could be due to reacceleration of particles in this region, caused by interaction with newer, more energetic particles from the outburst that created the NML. The new material catches up to the old lobes as it has a less dense medium to travel through, as shown in numerical simulations by \citet{clarke1991}. Finally, the spectrum steepens again at the end of the northern hook (point B in Fig.~\ref{tom1north}); this could indicate an older population of particles that has not experienced reacceleration, possibly because the central engine has precessed and the subsequent outbursts are not directed toward this region. The spectral variations we observe in Cen~A, however, are relatively subtle. In most cases $\Delta\alpha^{1400}_{118}<0.1$, while in Her~A $\Delta\alpha^{4800}_{1300}\approx0.8$ between the bridge and the jet and ring features \citep{HerA}. While some of the difference in magnitude of the spectral variation can be attributed to the lower frequency of the Cen~A spectral-index measurement, it is possible that variations in the magnetic field strength or the initial jet spectrum may be enough to cause such variations without the need for reacceleration.

The structure of Cen~A is complicated by the bending of the lobes, which is probably due to both the interaction with the ambient medium \citep{morganti}, at least for the northern lobe, and the precession or reorientation of the central engine over time \citep{haynes}. The radio morphology of Cen~A at small, intermediate and large scales supports the multiple-outburst scenario described above, where there appear to have been three epochs of activity that formed the inner, middle and outer lobes, respectively. However, the detailed timing of these outbursts of activity is difficult to determine and remains unclear. \citet{hardcastle2009} calculate spectral ages of $\sim$30 Myr for the outer lobes of Cen~A and \citet{haynes} calculate a precession rate for the central engine of $1\times10^{-5}$ degrees/yr from the stellar ages of observed B supergiants near the NML \citep{osmer}. This implies that over the time of formation of the outer lobes, there were several precessional periods of the central engine and also time for several outbursts of activity on similar scales to that which formed the inner lobes, which have an estimated spectral age of $\sim5$ Myr \citep{burns1983}. In a separate analysis, \citet{wykes} estimate much greater ages for the giant lobes of 440 to 645~Myr and 560~Myr (for both lobes) based on dynamical and buoyancy arguments respectively, which would allow time for many more outbursts of activity and precessional periods. \citet{wykes}, however, recognise that their age estimates differ from ages derived by radio spectral index information by a large amount and we note that the arguments they base their calculations on are known to overestimate lobe ages in radio galaxies as detailed in \citet{mcnamara}.



A problem with the multiple-outburst scenario is the apparent non-existence of a southern counterpart to the NML. Our 118~MHz map and spectral-tomography results do however indicate the existence of a steeper spectrum component along the southern lobe (see Figs \ref{tom2south} and \ref{SML}) at a location that is at approximately the correct position angle to be associated with the same period of activity as the NML. Whether this is evidence for the existence of a faint, steep spectrum feature associated with a `southern middle lobe' will be investigated through future observations with the full 128-tile MWA, which will provide higher angular resolution and increased sensitivity to such diffuse radio structures.

\subsection{Comparison to gamma-ray data}

\citet{yang} model the SEDs of the northern and southern lobes of Cen~A using gamma-ray observations from Fermi-LAT and radio data from ground-based telescopes (see \citealt{hardcastle2009}) and WMAP \citep{wmap2009,wmap2011}. They present SEDs from inverse-Compton models, which predict the flux density of the radio emission at low frequencies, integrated over the entire gamma-ray emitting region of the lobes.  Figs~8 and 9 of \citet{yang} show SEDs for two values of the model age of the outer lobes, $t$, selected based on the approximate lower limit of the lobe ages of $10^7$ years, set by dynamical arguments and the modelling of \citet{hardcastle2009}, and an upper limit of $10^8$ years, supported by the GeV data in the \citet{yang} paper itself. We determine the predicted 2-point spectral index between 118~MHz and 1.4~GHz for both of the \citet{yang} models. For a lobe age of 10~Myr the predicted spectral index in both the northern and southern lobes is approximately $-0.62$ and for a lobe age of 80~Myr the predicted spectral index in both lobes is $-0.57$. Our values for the average spectral indices of the lobes as derived from the T-T plots (Fig.~\ref{tt2}) of $\alpha_{1,2} = -0.66 \pm 0.01{\rm (stat)}^{+0.10_{\rm sys}}_{-0.08_{\rm sys}}$ in the northern lobe and $\alpha_{4} = -0.63 \pm 0.01{\rm (stat)}^{+0.10_{\rm sys}}_{-0.08_{\rm sys}}$ and $\alpha_{5} = -0.65 \pm 0.01{\rm (stat)}^{+0.10_{\rm sys}}_{-0.08_{\rm sys}}$ in the southern lobe, are consistent with both of these predictions. The difference between the predicted spectral indices is too small for us to distinguish between the lobe ages from our measured spectral indices and in any case the modelling of \citet{yang} can only be considered a first-order approximation to the true physical conditions. Assumptions such as a single age and spectral index for the lobes, a constant magnetic field and constant particle densities have been shown to be untrue (see e.g., \citealt{junkes,feain2011}, this work) therefore more detailed theoretical models are needed to take full advantage of current and future observational data.



\section{Conclusion}

We have imaged the entirety of Cen~A and a large surrounding field at 118~MHz with the MWA~32T. The wide field of view and good uv-coverage of the MWA~32T allow us to map the giant lobes at a reasonable angular resolution and to reconstruct their large-scale structure. We have investigated the spatially-resolved spectral properties of the giant lobes by comparing the MWA map at 118~MHz to Parkes data at 1.4~GHz. Through the use of T-T plots and spectral tomography, we find that the spectral index of the lobes between 118~MHz and 1.4~GHz has a spatially-averaged value consistent with measurements at other frequencies, but that there is a complex spatial distribution of spectral index across the lobes. The morphology and spectral-index distribution of the lobes supports a scenario of multiple outbursts from a precessing central engine, and we find tentative evidence for the existence of a faint, steep spectrum southern counterpart to the northern middle lobe. We find that our results also agree well with the inverse-Compton modelling of gamma-ray and radio data. Future observations with the complete 128-tile MWA will provide a more detailed picture of the low-frequency properties of Cen~A, which can be used to constrain more complex theoretical models. 


\section*{Acknowledgments}

This scientific work makes use of the Murchison Radio-astronomy Observatory, operated by CSIRO. We acknowledge the Wajarri Yamatji people as the traditional owners of the Observatory site. Support for the MWA comes from the U.S. National Science Foundation (grants AST-0457585, PHY-0835713, CAREER-0847753, and AST-0908884), the Australian Research Council (LIEF grants LE0775621 and LE0882938), the U.S. Air Force Office of Scientific Research (grant FA9550-0510247), and the Centre for All-sky Astrophysics (an Australian Research Council Centre of Excellence funded by grant CE110001020). Support is also provided by the Smithsonian Astrophysical Observatory, the MIT School of Science, the Raman Research Institute, the Australian National University, and the Victoria University of Wellington (via grant MED-E1799 from the New Zealand Ministry of Economic Development and an IBM Shared University Research Grant). The Australian Federal government provides additional support via the Commonwealth Scientific and Industrial Research Organisation (CSIRO), National Collaborative Research Infrastructure Strategy, Education Investment Fund, and the Australia India Strategic Research Fund, and Astronomy Australia Limited, under contract to Curtin University. We acknowledge the iVEC Petabyte Data Store, the Initiative in Innovative Computing and the CUDA Center for Excellence sponsored by NVIDIA at Harvard University, and the International Centre for Radio Astronomy Research (ICRAR), a Joint Venture of Curtin University and The University of Western Australia, funded by the Western Australian State government. We would like to thank P. Leahy for many constructive comments and suggestions that served to improve the manuscript.

\bsp

\label{lastpage}


\begin{thebibliography}{99}

\bibitem[\protect\citeauthoryear{Abdo et al.}{2010}]{abdo} Abdo, A. A., Ackermann, M., Ajello, M., et al., 2010, Sci, 328, 725
\bibitem[\protect\citeauthoryear{Alvarez et al.}{1997}]{alvarez1997} Alvarez, H., Aparici, J., May, J., Olmos, F., 1997, A\&AS, 124, 315
\bibitem[\protect\citeauthoryear{Alvarez et al.}{2000}]{alvarez} Alvarez, H., Aparici, J., May, J.,  Reich, P., 2000, A\&A, 355, 863
\bibitem[\protect\citeauthoryear{Atwood et al.}{2009}]{fermiLAT} Atwood, W. B., Abdo, A. A., Ackermann, M., et al., 2009, ApJ, 697, 1071
\bibitem[\protect\citeauthoryear{Baars et al.}{1977}]{baars} Baars, J. W. M., Genzel, R., Pauliny-Toth, I. I. K., et al., 1977, A\&A, 61, 99
\bibitem[\protect\citeauthoryear{Begelman et al.}{1984}]{begelman} Begelman, M. C., Blandford, R. D., Rees, M. J., 1984, Reviews of Modern Physics, 56, 255
\bibitem[\protect\citeauthoryear{Bernardi et al.}{2009}]{bernardi} Bernardi, G., de Bruyn, A. G., Brentjens, M. A., et al., 2009, A\&A, 500, 965
\bibitem[\protect\citeauthoryear{Bernardi et al.}{2013}]{bernardi2013} Bernardi, G., Greenhill, L. J., Mitchell, D. A., et al., 2013, ApJ, 771, 105
\bibitem[\protect\citeauthoryear{Boggs \& Rogers}{1990}]{ODR} Boggs, P. T., Rogers, J. E., 1990, Contemporary Mathematics, 112, 186
\bibitem[\protect\citeauthoryear{Bolton}{1948}]{bolton} Bolton, J. G., 1948, Nature, 162, 141
\bibitem[\protect\citeauthoryear{Bowman et al.}{2013}]{bowman2013} Bowman, J. D., Cairns, I., Kaplan, D. L., et al, 2013, PASA, 30, 31
\bibitem[\protect\citeauthoryear{Burns}{1986}]{burns} Burns, J. O., 1986, Canadian Journal of Physics, 64, 373
\bibitem[\protect\citeauthoryear{Burns et al.}{1983}]{burns1983} Burns, J. O., Feigelson, E. D., Schreier, E. J., 1983, ApJ, 273, 128
\bibitem[\protect\citeauthoryear{Clarke et al.}{1992}]{vla} Clarke, D. A., Burns, J. O., Norman, M. L., 1992, ApJ, 395, 444
\bibitem[\protect\citeauthoryear{Clarke \& Burns}{1991}]{clarke1991} Clarke, D. A., Burns, J. O., 1991, ApJ, 369, 308
\bibitem[\protect\citeauthoryear{Combi \& Romero}{1997}]{combi} Combi, J. A., Romero, G. E., 1997, A\&AS, 121, 11
\bibitem[\protect\citeauthoryear{Cooper et al.}{1965}]{cooper1965} Cooper, B. F. C.; Price, R. M.; Cole, D. J., 1965,  Australian Journal of Physics, 18, 589
\bibitem[\protect\citeauthoryear{Cornwell et al.}{2008}]{wproj} Cornwell, T. J., Golap, K. \& Bhatnagar, S., 2008, IEEE Journal of Selected Topics in Signal Processing, 2, 647
\bibitem[\protect\citeauthoryear{DeLaney et al.}{2002}]{delaney} DeLaney, T., Koralesky, B., Rudnick, L., Dickel, J. R., 2002, ApJ, 580, 914
\bibitem[\protect\citeauthoryear{de Oliveira-Costa et al.}{2008}]{deoliveiracosta2008} de Oliveira-Costa, A., Tegmark, M., Gaensler, B. M., et al., 2008, MNRAS, 388, 247
\bibitem[\protect\citeauthoryear{Ellis \& Hamilton}{1966}]{ellis} Ellis, G. R. A., Hamilton, P. A, 1966, ApJ, 143, 227
\bibitem[\protect\citeauthoryear{Fanaroff \& Riley}{1974}]{FR} Fanaroff, B. L., Riley, J. M., 1974, MNRAS, 167, 31
\bibitem[\protect\citeauthoryear{Feain et al.}{2011}]{feain2011} Feain, I. J., Cornwell, T. J., Ekers, R. D., et al., 2011, ApJ, 740, 17
\bibitem[\protect\citeauthoryear{Furlanetto et al.}{2006}]{furlanetto2006} Furlanetto, S. R., Oh, S. P., Briggs, F. H., et al., 2006, Physics Reports, 433, 4-6
\bibitem[\protect\citeauthoryear{Gaensler \& Wallace}{2003}]{gaenslerwallace} Gaensler, B. M., Wallace, B. J., 2003, ApJ, 594, 326
\bibitem[\protect\citeauthoryear{Gizani \& Leahy}{2003}]{HerA} Gizani, N. A. B., Leahy, J. P., 2003, MNRAS, 342, 399
\bibitem[\protect\citeauthoryear{Giacintucci et al.}{2007}]{gia} Giacintucci, S., Venturi, T., Murgia, M., et al., 2007, A\&A, 476, 99
\bibitem[\protect\citeauthoryear{Gopal-Krishna \& Saripalli}{1984}]{gopal1984} Gopal-Krishna, Saripalli, L., 1984, A\&A, 141, 61
\bibitem[\protect\citeauthoryear{Gopal-Krishna \& Wiita}{2010}]{gopal2010} Gopal-Krishna, Wiita, P. J., 2010, NewA, 15, 96
\bibitem[\protect\citeauthoryear{Hamilton \& Haynes}{1968}]{hamilton} Hamilton, P. A. \& Haynes, R. F., 1968,  Australian Journal of Physics, 21, 895
\bibitem[\protect\citeauthoryear{Hardcastle et al.}{2009}]{hardcastle2009} Hardcastle, M. J., Cheung, C. C., Feain, I. J., Stawarz, \L., 2009, MNRAS, 393, 1041
\bibitem[\protect\citeauthoryear{Harris et al.}{2010}]{harris} Harris, G. L. H., Rejkuba M., Harris W. E., 2010,  PASA, 27, 457
\bibitem[\protect\citeauthoryear{Haslam et al.}{1982}]{haslam} Haslam, C. G., Salter C. J., Stoffel H., Wilson W. E., 1982, A\&AS, 47, 1
\bibitem[\protect\citeauthoryear{Haynes et al.}{1983}]{haynes} Haynes, R. F., Cannon, R. D., Ekers, R. D., 1983, PASA, 5, 241
\bibitem[\protect\citeauthoryear{Hinshaw et al.}{2009}]{wmap2009} Hinshaw, G., Weiland, J. L., Hill, R. S., et al, 2009, ApJS, 180, 225
\bibitem[\protect\citeauthoryear{Israel}{1998}]{israel} Israel, F. P., 1998, A\&ARv, 8, 237
\bibitem[\protect\citeauthoryear{Jacobs et al.}{2013}]{jacobs2013} Jacobs, D. C., Bowman, J., Aguirre, J. E., 2013, ApJ, 769, 5
\bibitem[\protect\citeauthoryear{Jamrozy et al.}{2009}]{jamrozy} Jamrozy, M., Konar, C., Saikia, D. J., Machalski, J., 2009, in ASP Conf. Ser. 36, The Low-Frequency Radio Universe, ed. D. J. Saikia, D. A. Green, Y. Gupta, \& T. Venturi (San Francisco, CA: ASP), 137
\bibitem[\protect\citeauthoryear{Junkes et al.}{1993}]{junkes} Junkes, N., Haynes, R. F., Harnett, J. I., Jauncey, D. L., 1993, A\&A, 269, 29
\bibitem[\protect\citeauthoryear{Katz-Stone \& Rudnick}{1997}]{katz-stone} Katz-Stone, D. M., Rudnick, L., 1997, ApJ, 488, 146
\bibitem[\protect\citeauthoryear{Komatsu et al.}{2011}]{wmap2011} Komatsu, E., Smith, K. M., Dunkley, J., et al., 2011, ApJS, 192, 18
\bibitem[\protect\citeauthoryear{Large et al.}{1981}]{pks1328-410} Large, M. I., Mills, B. Y., Little, A. G., et al.,1981, MNRAS, 194, 693
\bibitem[\protect\citeauthoryear{Leahy et al.}{1986}]{leahy} Leahy, J. P., Pooley, G. G., Riley, J. M., 1986, MNRAS, 222, 753
\bibitem[\protect\citeauthoryear{Lonsdale et al.}{2009}]{lonsdale} Lonsdale, C. J., Cappallo, R. J., Morales, M. F., et al, 2009, in Proc. IEEE, 97, 8
\bibitem[\protect\citeauthoryear{Ma et al.}{1998}]{ICRF} Ma, C., Arias, E. F., Eubanks, T. M., et al., 1998, AJ, 116, 516
\bibitem[\protect\citeauthoryear{Malin et al.}{1983}]{malin} Malin, D. F., Quinn, P. J., Graham, J. A., 1983, ApJ, 272, 5
\bibitem[\protect\citeauthoryear{Mauch et al.}{2003}]{sumss} Mauch, T., Murphy, T., Buttery, H. J., et al., 2003, MNRAS, 342, 1117
\bibitem[\protect\citeauthoryear{McNamara \& Nulsen}{2007}]{mcnamara} McNamara, B. R. \& Nulsen, P. E. J., 2007, ARA\&A, 45, 117
\bibitem[\protect\citeauthoryear{Morales \& Wyithe}{2010}]{moraleswyithe2010} Morales, M. F., Wyithe, J. S. B., 2010, ARA\&A, 48, 127
\bibitem[\protect\citeauthoryear{Morganti et al.}{1999}]{morganti} Morganti, R., Killeen, N. E. B., Ekers, R. D., Oosterloo, T. A., MNRAS, 307, 750
\bibitem[\protect\citeauthoryear{Norman et al.}{1988}]{wats} Norman, M. L., Burns, J. O., Sulkanen, M. E., 1988, Nature, 335, 146
\bibitem[\protect\citeauthoryear{Offringa et al.}{2010}]{offringa2010} Offringa, A. R., de Bruyn, A. G., Biehl, M., et al., 2010, MNRAS, 405, 155
\bibitem[\protect\citeauthoryear{Offringa et al.}{2012}]{offringa2012} Offringa, A. R., van de Gronde, J. J., Roerdink, J. B. T. M., et al., 2012, A\&A, 539, 95
\bibitem[\protect\citeauthoryear{Osmer}{1978}]{osmer} Osmer, P. S., 1978, ApJ, 226, 79
\bibitem[\protect\citeauthoryear{O'Sullivan et al.}{2013}]{osullivan} O'Sullivan, S. P., Feain, I. J., McClure-Griffiths, N. M., et al., 2013, ApJ, 764, 162
\bibitem[\protect\citeauthoryear{Parsons et al.}{2010}]{PAPER} Parsons, A. R., Backer, D. C., Foster, G. S., et al., 2010, AJ, 139, 1468
\bibitem[\protect\citeauthoryear{Pritchard \& Loeb}{2012}]{pritchard_loeb} Pritchard, J. R. \& Loeb, A., 2012, Reports on Progress in Physics, 75, 086901
\bibitem[\protect\citeauthoryear{Reynolds}{1994}]{reynolds1994} Reynolds, J. E. 1994, ATNF Technical Memo Series, 39.3/040
\bibitem[\protect\citeauthoryear{Saxton et al.}{2001}]{saxton} Saxton, C. J., Sutherland, R. S., Bicknell, G. V., 2001, ApJ, 563, 103
\bibitem[\protect\citeauthoryear{Schoenmakers et al.}{2000}]{schoenmakers2000} Schoenmakers, A. P., de Bruyn, A. G., R\"{o}ttgering, H. J. A., et al., 2000, MNRAS, 315, 371
\bibitem[\protect\citeauthoryear{Schreier et al.}{1981}]{schreier} Schreier, E. J., Burns, J. O., Feigelson, E. D., 1981, ApJ, 251, 523
\bibitem[\protect\citeauthoryear{Shain}{1959}]{shain} Shain, C. A., 1959, IAUS, 9, 328
\bibitem[\protect\citeauthoryear{Sheridan}{1958}]{sheridan} Sheridan, K. V., 1958, Australian Journal of Physics, 11, 400
\bibitem[\protect\citeauthoryear{Schilizzi \& McAdam}{1975}]{schilizzi} Schilizzi, R. T., McAdam, W. B., 1975, Royal Astronomical Society Memoirs, 79, 1
\bibitem[\protect\citeauthoryear{Schiminovich et al.}{1994}]{HIshells} Schiminovich, D., van Gorkom, J. H., van der Hulst, J. M., et al., 1994, ApJL, 423, 101
\bibitem[\protect\citeauthoryear{Slee}{1977}]{culgoora} Slee, O. B., 1977, Australian Journal of Physics Astrophysical Supplement, 43, 1
\bibitem[\protect\citeauthoryear{Slee}{1995}]{culgoora1995} Slee, O. B., 1995, Australian Journal of Physics, 48, 143
\bibitem[\protect\citeauthoryear{Stefan et al.}{2013}]{stefan} Stefan, I. I., Carilli, C. L., Green, D. A., et al., 2013, MNRAS, 432, 1285
\bibitem[\protect\citeauthoryear{Stickel et al.}{2004}]{stickel} Stickel, M., van der Hulst, J. M., van Gorkom, et al., 2004, A\&A, 415, 95
\bibitem[\protect\citeauthoryear{Subrahmanyan et al.}{1996}]{ravi}Subrahmanyan, R., Saripalli, L., Hunstead, R. W., 1996, MNRAS, 279, 257
\bibitem[\protect\citeauthoryear{Tingay et al.}{2013}]{tingay} Tingay, S. J., Goeke, R., Bowman, J. D., et al., 2013, PASA, 30, 7
\bibitem[\protect\citeauthoryear{Turtle et al.}{1962}]{turtle} Turtle, A. J., Pugh, J. F., Kenderdine, S., Pauliny-Toth, I. I. K., 1962, MNRAS, 124, 297
\bibitem[\protect\citeauthoryear{van Haarlem et al.}{2013}]{LOFAR} van Haarlem, M. P., Wise, M. W., Gunst, A. W., et al., 2013, A\&A in press (arXiv:1305.3550)
\bibitem[\protect\citeauthoryear{Williams et al.}{2012}]{williams} Williams, C. L., Hewitt, J. N., Levine, A. M., et al.,  2012, ApJ, 755, 47
\bibitem[\protect\citeauthoryear{Wykes et al.}{2013}]{wykes} Wykes, S., Croston, J. H., Hardcastle, M. J., et al., 2013, A\&A in press (arXiv:1305.2761)
\bibitem[\protect\citeauthoryear{Yang et al.}{2012}]{yang} Yang, R. Z., Sahakyan, N., de Ona Wilhelmi, E., et al., 2012, A\&A, 542, 19 
\end{thebibliography}
\end{document}